\documentclass[aps,twocolumn,amssymb,showpacs,floatfix]{revtex4-1}
\usepackage{graphicx}
\usepackage{multirow}
\usepackage[breaklinks,colorlinks,urlcolor=blue,citecolor=blue]{hyperref}
\usepackage{mathrsfs}
\usepackage{amsmath}
\usepackage{soul}
\usepackage{color}
\usepackage[normalem]{ulem}
\usepackage[makeroom]{cancel}
\usepackage{amssymb}
\usepackage{float}
\graphicspath{{./figs}}

\begin{document}

\title{Hot and Dense Matter Equation of State Probability
  Distributions \\ for Astrophysical Simulations}
\author{Xingfu Du$^{1}$}
\author{Andrew W. Steiner$^{1,2}$}
\author{Jeremy W. Holt$^{3}$}
\affiliation{$^{1}$Department of Physics and Astronomy, University of
  Tennessee, Knoxville, TN 37996, USA}
\affiliation{$^{2}$Physics Division, Oak Ridge National Laboratory, Oak
  Ridge, TN 37831, USA}
\affiliation{$^{3}$Cyclotron Institute and Department of Physics and
  Astronomy, Texas A\&M University, College Station, Texas 77843, USA}

\begin{abstract}
  We add an ensemble of nuclei to the equation of state for
  homogeneous nucleonic matter to generate a new set of models
  suitable for astrophysical simulations of core-collapse supernovae
  and neutron star mergers. We implement empirical constraints from
  (i) nuclear mass measurements, (ii) proton-proton scattering phase
  shifts, and (iii) neutron star observations. Our model is also
  guided by microscopic many-body theory calculations based on
  realistic nuclear forces, including the zero-temperature neutron
  matter equation of state from quantum Monte Carlo simulations and
  thermal contributions to the free energy from finite-temperature
  many-body perturbation theory. We ensure that the parameters of our
  model can be varied while preserving thermodynamic consistency and
  the connection to experimental or observational data, thus providing
  a probability distribution of the astrophysical hot and dense matter
  equation of state. We compare our results with those obtained from
  other available equations of state. While our probability
  distributions indeed represent a large number of possible equations
  of state, we cannot yet claim to have fully explored all of the
  uncertainties, especially with regard to the structure of nuclei in
  the hot and dense medium. 
\end{abstract}


\maketitle

\section{Introduction}
The equation of state (EOS) of nuclear matter is a central microscopic
input for the simulation of core-collapse supernovae and neutron star
mergers. In a supernova, the nuclear incompressibility generated from
Fermi degeneracy pressure and short-range nuclear forces is essential
in providing the pressure support which causes the infalling shockwave
to ``bounce'' and propel the mantle off the protoneutron star
underneath~\cite{Bethe79}. In a neutron star merger, the EOS
determines the compactness of the two stars, which in turn determines
the amount of r-process material ejected in a
merger~\cite{Sekiguchi15dm}, the properties of the kilonova emission
\cite{Kasen2017}, and features of the late-inspiral gravitational wave
emissions (e.g. see Ref.~\cite{Hinderer08}). The EOS also determines
the lifetime and final fate of the merger remnant
\cite{bauswein17,margalit17,radice18,rezzolla18,ruiz18,shibata17}
through the relationship between the EOS and the neutron star maximum
mass.

Since weak equilibrium is not fully achieved in the short dynamical
timescale of either a supernova explosion or a neutron star merger,
there are at least three relevant quantities for describing the
composition of dense matter: the number density of baryons $n_B$, the
electron fraction $Y_e$, and the temperature $T$. Muons, pions, and
strangeness-containing hadrons may introduce additional complexity,
but as a minimal model we neglect these more exotic degrees of freedom
in the present work. Simulations of supernovae or mergers which employ
realistic EOSs often use tabulations that span baryon number densities
$n_B \sim 10^7 - 10^{15}$\,g/cm$^3$, electron fractions $Y_e \sim
0.1-0.6$, and temperatures $T \sim 0 - 100$\,MeV.

EOSs for core-collapse supernovae were first developed by Lattimer and
Swesty~\cite{Lattimer91}, who employed three different
non-relativistic Skyrme effective interactions and the single-nucleus
approximation to account for the presence of heavy nuclei in a gas of
unbound nucleons. A second set of EOS tables was developed H. Shen et
al.~\cite{Shen98} (also using the single-nucleus approximation), which
was based on the NL3 relativistic mean-field Lagrangian. While the
single-nucleus approximation is sufficient to describe the bulk
thermodynamics, it does not in general accurately describe the
composition~\cite{Burrows84,Hix03,Botvina05,OConnor07,Arcones08,Souza09b}
and the associated weak reaction rates. G. Shen et al.~\cite{Shen10}
constructed the first full table to go beyond the single-nucleus
approximation. Their work was based on a more modern relativistic
mean-field model, ``FSUGold''~\cite{ToddRutel05}, and goes beyond the
single nucleus approximation to include a full distribution of nuclei
in nuclear statistical equilibrium (NSE). Alternative formalisms were
developed by Furusawa et al.~\cite{Furusawa_2011} and Hempel et
al.~\cite{Hempel10,Hempel12}, which resulted in EOS tables built upon
several nucleon-nucleon interactions, including FSUGold,
DD2~\cite{Typel10}, IUFSU~\cite{Fattoyev10}, SFHo~\cite{Steiner13cs}
and SFHx~\cite{Steiner13cs}. More recently, several EOSs have been
added to the CompOSE (CompStar Online Supernovae Equations of State)
database~\cite{Typel13cc}, including an EOS with
hyperons~\cite{Banik14}. Recent EOS tables with a similar goal of
matching observational and experimental constraints have been released
by Schneider et al.~\cite{Schneider19es,Schneider19ae}.

The basic paradigm under which most EOS tables are constructed is to
compute the thermodynamic quantities based on a single model of the
nucleon-nucleon interaction. However, this paradigm fails when one
wants to perform {\em{uncertainty quantification}}. There is currently
no model for the nucleon-nucleon interaction and the accompanying EOS
which (i) faithfully describes matter in all of the density and
temperature regimes which are relevant for supernovae and mergers and
(ii) allows one to vary a set of parameters in such a way as to
explore the uncertainties in the EOS without spoiling agreement with
experiments or observations. For example, Skyrme~\cite{Skyrme59}
models are often used to describe dense matter for the purposes of EOS
tables, but often fail to describe low-density matter as described by
the virial expansion or nuclear effective field theory
\cite{Kruger13}. Even when a Skyrme effective interaction does happen
to match model-independent properties of the EOS at low-densities, it
does so at the cost of suppressing the uncertainties in matter at
higher densities and introducing unphysical correlations between
matter in the two density regimes.

Future work on the nucleon-nucleon interaction and the equation of
state may eventually resolve some of these issues. In the meantime, a
different approach is required to ensure that simulations can quantify
the uncertainties in the EOS without over- or underconstraining the
EOS. Based on our previous work in Ref.~\cite{Du19hd}, we construct a
phenomenological description of the free energy for hot and dense
stellar matter which is able to (i) faithfully describe nuclear matter
under conditions that are probed by nuclear experiments and
observations of neutron stars and (ii) provides parameters which allow
one to (at least partially) quantify the uncertainties which result
from our imperfect knowledge of the nucleon-nucleon interaction. We
add nuclei to the EOS of homogeneous nuclear matter described in
Ref.~\cite{Du19hd} and show that our results compare well with other
EOS tables which are available.

\section{Method}

\subsection{Basic Formalism}
We use the formalism developed in Ref.~\cite{Hempel10} to describe
nucleons in thermodynamic equilbrium with a distribution of nuclei.
Neutrons, protons, $\alpha$ particles, deuterons, tritons, $^4 \rm
Li$, and $^3 \rm He$ are treated separately to more easily describe
the neutrino opacities near the neutrinosphere~\cite{Horowitz12cc}.
The Helmholtz free energy density can be written as
\begin{equation}
  f(n_n, n_p, \{n_i\}, T) = f_{np} + \sum_{i} f_i +
  f_{\mathrm{Coul}} + f_e,
\end{equation}
where $n_n$ and $n_p$ are the free neutron and proton number
densities, $n_i$ is the number density of nucleus $i$, and
$f_{\mathrm{Coul}}$ denotes the Coulomb free energy described in more
detail below. We take $\hbar=c=k_B=1$. Baryon number conservation and
global charge neutrality imply two constraints, which we write as
\begin{eqnarray}
  \label{eq:nbye}
  n_B &=& n_n + n_p + \sum_i n_i A_i \nonumber \\
  n_B Y_e &=& n_e = n_p + \sum_i n_i Z_i \, .
  \label{eq:cons}
\end{eqnarray}

The free energy density of nucleons outside the nucleus, denoted
$f_{\mathrm{Hom}}$, is based on the homogenous nucleonic matter EOS
from Ref.~\cite{Du19hd} (see discussion below). (See also
Ref.~\cite{Huth21} for an alternative EOS for homogeneous nucleonic
matter.) We include an excluded volume correction (which is only
turned on between nucleons and nuclei), to correct for the fact that
the volume available to the nucleons is reduced by the nuclei. We
denote the volume available to nucleons as $V^{\prime} \equiv V-
\sum_{i} N_i V_i$, where $N_i \equiv n_i V $ is the number of nuclei
of type $i$ in the volume $V$, $V_i \equiv A_i/n_0$ is the volume
occupied by one nucleus of type $i$, and $n_0$ is the saturation
density of symmetric nuclear matter, $0.16~\mathrm{fm}^{-3}$. The
volume fraction that free nucleons explore is $\xi \equiv
V^{\prime}/V=1-\sum_{i} A_i n_i/ n_0$. Thus,
\begin{equation}
  f_{np} = \xi f^{\prime}_{\mathrm{Hom}} (n^{\prime}_n,n^{\prime}_p,T)
  \, .
\end{equation}
The $n^{\prime}_n$ and $n^{\prime}_p$ are local densities and are
defined by $n^{\prime}_n \equiv N_n/V^{\prime}= n_n/\xi$,
$n^{\prime}_p \equiv N_p/V^{\prime}= n_p/\xi$ separately. We ignore
rest mass contribution here and put a tilde on top when it is added
back.

The free energy density of the light nuclei and heavy nucleus are
treated as classical Boltzmann particles:
\begin{equation}
  f_{i} = - n_i T \left[ \ln \left( \frac{\Omega_i \bar{V}}{N_i
      \lambda_i^3} \right)+1
    \right],
\end{equation}
where $\lambda_i$ is the thermal wavelength
\begin{equation}
  \lambda_i = \left( \frac{2 \pi}{ m_i T}\right)^{1/2} \,
\end{equation}
and $\bar{V} \equiv \kappa V$ is the volume fraction explorable
to the nucleus of type $i$, with $\kappa \equiv 1-n_B/n_0$.
The quantity $\Omega_i$ is the temperature-dependent partition
function. The prescription we use follows from
Refs.~\cite{Fowler78np,Shen10} and will be addressed in the next
section. Using
these definitions
\begin{equation}
  f_i = - n_i T
  \left[ \ln\left(\frac{\Omega_i}{n_i \lambda_i^3} \right)+1 \right]
  - n_i T \ln \kappa.
\end{equation}
One can also rewrite $\xi$ in terms of $\kappa$ and the nucleon
densities
\begin{equation}
  \xi=\kappa+(n_n+n_p)/n_0 = \kappa \left( 1 -
  \frac{n_n^{\prime}}{n_0} - \frac{n_p^{\prime}}{n_0} \right)^{-1} \, .
  \label{eq:xi2}
\end{equation}

The Coulomb energy in the Wigner-Seitz cell is~\cite{Baym71}
\begin{equation}
  E_i^{\mathrm{Coul}} =- \frac{3}{5} \frac{Z_i^2 \alpha} {R_i}
  \left(\frac{3}{2} x_i - \frac{1}{2}x_i^3 \right),
\end{equation}
where
\begin{equation}
  x_i \equiv \left(\frac{n_B Y_e}{n_0} \frac{A_i}{Z_i} \right)^{1/3}
  = \frac{R_i}{R_{\mathrm{WS},i}},
\end{equation}
where $R_i^3 = (3 A_i)/(4 \pi n_0)$ is the nuclear radius and the
size of the Wigner-Seitz cell, $R_{\mathrm{WS}}$, is given by
\begin{equation}
  Z_i = \frac{4 \pi}{3} R_{\mathrm{WS},i}^3 n_B Y_e \, .
\end{equation}
The radius of nuclei is constrained by $R_i \leq R_{\mathrm{ws}, i}$
which limits $x_i \leq 1$.

We take into account all charged particles here except protons, the
advantage is the Coulomb energy is merely a function of charge and
atomic number given $n_B$ and $Y_e$. After applying charge
neutrality, the total free energy density becomes
\begin{eqnarray}
  f(n_n,n_p,\{n_i\},T) &=& \xi f_{\mathrm{Hom}}
  (n^{\prime}_n,n^{\prime}_p,T)
  \nonumber \\ && -
  \sum_{i} n_i T \left\{ \left[ \ln \left(\frac{\Omega_i}{n_i \lambda_i^3}
    \right)+1 \right] +
  \ln \kappa \right\}
  \nonumber \\ && +
  \sum_{i} n_i E^{\mathrm{Coul}}_i + f_e(n_B Y_e).
  \label{eq:f_full}
\end{eqnarray}

\subsection{Homogeneous matter}

We slightly modify the EOS of homogeneous matter from
Ref.~\cite{Du19hd} to no longer enforce a quadratic expansion for the
isospin-asymmetry dependence of the 
finite-temperature contributions.
The free energy is separated into a contribution from
the virial expansion and a contribution from degenerate matter:
\begin{eqnarray}
  f_{\mathrm{Hom}}(n_B,x_p,T) &=& f_{\mathrm{virial}}(n_B,x_p,T) g \nonumber
  \\ && + f_{\mathrm{deg}}(n_B,x_p,T) (1-g) \, .
\end{eqnarray}
The free energy density for degenerate matter is
\begin{eqnarray}
  f_{\mathrm{deg}}(n_B,x_p,T) &=&
  f_{\mathrm{Skyrme}}(n_B,x_p=1/2,T=0) \qquad
  \nonumber \\
  && \hspace{-.2in}+ \delta^2 {\varepsilon}_{\mathrm{sym}}(n_B)
  + \Delta f_{\mathrm{hot}}(n_B,x_p,T).
\end{eqnarray}
Based on the work in Ref.~\cite{Zhang18nd}, we use the Skyrme model
labeled SK$\chi$m$^{*}$ to compute $\Delta f_{\mathrm{hot}}$. This Skryme
model was fitted to the equation of state of asymmetric nuclear matter
\cite{sammarruca15,Wellenhofer15to} 
calculated from several realistic chiral two- and three-body forces 
as well as consistent nucleon isoscalar and isovector 
effective masses derived from the nucleon self
energy \cite{holt13prcb,holt16prc}. In particular, the description of nuclear
matter thermal properties relies on accurately modeling the nucleon effective 
mass, which is proportional to the density of states near the Fermi surface
and hence the temperature 
dependence of the entropy \cite{Constantinou14tp,rrapaj16}. The
derivatives of the degenerate free energy density are
\begin{eqnarray}
  \frac{\partial f_{\mathrm{deg}}}{\partial n_n}  &=&
  \frac{1}{2} \mu_{n,\mathrm{Skyrme}}(n_B,x_p=1/2,T=0)
  \nonumber \\
  &&+ \frac{1}{2} \mu_{p,\mathrm{Skyrme}}(n_B,x_p=1/2,T=0)
  \nonumber \\ && + \delta^2
  \frac{\partial \varepsilon_{\mathrm{sym}}}{\partial n_B} +
  \frac{2 \delta (1-\delta)}{n_B} \varepsilon_{\mathrm{sym}}
  \nonumber \\
  &&  + \Delta \mu_{n,\mathrm{hot}}(n_B,x_p,T),
\end{eqnarray}
\begin{eqnarray}
  \frac{\partial f_{\mathrm{deg}}}{\partial n_p}  &=&
  \frac{1}{2} \mu_{p,\mathrm{Skyrme}}(n_B,x_p=1/2,T=0)
  \nonumber \\ &&  +\frac{1}{2} \mu_{n,\mathrm{Skyrme}}(n_B,x_p=1/2,T=0)
  \nonumber \\ && + \delta^2
  \frac{\partial \varepsilon_{\mathrm{sym}}}{\partial n_B} -
  \frac{2 \delta \left( 1 + \delta \right)}{n_B} \varepsilon_{\mathrm{sym}}
  \nonumber \\ 
  &&  + \Delta \mu_{p,\mathrm{hot}}(n_B,x_p,T),
\end{eqnarray}
and
\begin{equation}
  \frac{\partial f_{\mathrm{deg}}}{\partial T} =
  -  s_{\mathrm{hot}} \left(n_B, x_p,T \right)
\end{equation}
where
\begin{eqnarray}
  \frac{\partial \varepsilon_{\mathrm{sym}}}{\partial n_B} &=&
  h^{\prime}(n_B) {\varepsilon}_{\mathrm{QMC}}(n_B)
  + h(n_B) {\varepsilon}^{\prime}_{\mathrm{QMC}}(n_B)
  \nonumber \\ &&
  - h^{\prime}(n_B) {\varepsilon}_{\mathrm{NS}}(n_B)
  + \left[1-h(n_B)\right] {\varepsilon}^{\prime}_{\mathrm{NS}}(n_B) +
  \nonumber \\ &&
  - \frac{1}{2} \left[\mu_{n,\mathrm{Skyrme}}(n_B,x_p=1/2,T=0) \right.
    \nonumber \\ &&
    + \left. \mu_{p,\mathrm{Skyrme}}(n_B,x_p=1/2,T=0) \right].
    \label{esymn}
\end{eqnarray}
In Eq.\ \eqref{esymn}, the auxiliary function $h^\prime$ is used to
interpolate between the pure neutron matter equation of state
${\varepsilon}_{\mathrm{QMC}}$ valid around normal nuclear densities
and the high-density equation of state ${\varepsilon}_{\mathrm{NS}}$
that may be constrained by neutron star observations. Note that
${\varepsilon}_{\mathrm{QMC}}$ is given by the quantum Monte
Carlo-inspired form
\begin{equation}
  {\varepsilon}_{\mathrm{QMC}} = n_B \left
  ( a \left (\frac{n_B}{n_0} \right)^\alpha  
+ b \left (\frac{n_B}{n_0} \right)^\beta \right ).
\label{qmcnm}
\end{equation}

\subsection{The Saha equations}

In order to fix the densities of the nuclei, we solve the equations
\begin{equation}
  \left( \frac{\partial f}{\partial n_i}
  \right)_{n_B,Y_e} =
  0 \, .
\end{equation}
Before we begin, it is useful to define
\begin{equation}
  f_{i,\mathrm{Cl}} = - n_i T
  \left[ \ln\left(\frac{\Omega_i}{n_i \lambda_i^3} \right)+1 \right]
\end{equation}
as the classical part of the nuclear free energy.

We can rewrite the full free energy from Eq.~\eqref{eq:f_full}
in terms of $n_B$ and $Y_e$
\begin{equation}
  f[n_n(n_B,Y_e,T),n_p(n_B,Y_e,T),\{n_i\},T] \, ,
\end{equation}
to re-express the derivative ($T$ is implicitly held
constant)
\begin{eqnarray}
  \left( \frac{\partial f}{\partial n_i}
  \right)_{n_B,Y_e} &=&
  \left( \frac{\partial f}{\partial n_i}
  \right)_{n_p,n_n} +
  \left( \frac{\partial f}{\partial n_n}
  \right)_{n_p,\{n_i\}}
  \left( \frac{\partial n_n}{\partial n_i}\right)_{n_B,Y_e}
  \nonumber \\ &&
  +
  \left( \frac{\partial f}{\partial n_p}
  \right)_{n_n,\{n_i\}}
  \left( \frac{\partial n_p}{\partial n_i}\right)_{n_B,Y_e}
  \nonumber \\
  &=& 
  \left( \frac{\partial f}{\partial n_i}
  \right)_{n_p,n_n} -
  \left( \frac{\partial f}{\partial n_n}
  \right)_{n_p,\{n_i\}} N_i
  \nonumber \\ && -
  \left( \frac{\partial f}{\partial n_p}
  \right)_{n_n,\{n_i\}} Z_i \, .
\end{eqnarray}
Thus, we obtain the Saha equations
\begin{equation}
  \mu_i = \mu_n N_i + \mu_p Z_i \, ,
\end{equation}
with the chemical potentials defined by
\begin{eqnarray}
  \mu_i &\equiv& \left( \frac{\partial f}{\partial n_i}
  \right)_{n_p,n_n,T}, \quad
  \mu_n \equiv \left( \frac{\partial f}{\partial n_n}
  \right)_{n_p,\{n_i\},T} , \quad \mathrm{and}
  \nonumber \\ 
  \mu_p &\equiv& \left( \frac{\partial f}{\partial n_p}
  \right)_{n_n,\{n_i\},T} \, .
  \label{eq:mu_defn}
\end{eqnarray}
We also define $\mu_e \equiv (\partial f_e)/(\partial n_e)$.
These chemical potentials can be written analytically.
For the nuclei, we define
\begin{equation}
  P^{\mathrm{Coul}}_i \equiv - n_i \frac{3}{5} \frac{Z_i^2 \alpha} {R_i}
  \left(\frac{1}{2} x_i - \frac{1}{2}x_i^3 \right) \, .
\end{equation}
as in Ref.~\cite{Hempel10} and this definition implies
\begin{equation}
  n_i \left( \frac{\partial E_i^{\mathrm{Coul}} }
  {\partial n_i} \right) _{n_n,n_p} =
  \frac{P_i^{\mathrm{Coul}} Z_i}{n_e} \, .
\end{equation}
For the nuclei, this definition gives
\begin{eqnarray}
  \mu_i &=& -\left(\frac{A_i}{n_0}\right) f_{\mathrm{Hom}} +
  \mu_{n,\mathrm{Hom}}
  \left(\frac{A_i n_n}{\xi n_0 }\right)
  + \mu_{p,\mathrm{Hom}}
  \left(\frac{A_i n_p}{\xi n_0 }\right)
  \nonumber \\ && +
  \mu_{i,\mathrm{cl}} - T \ln \kappa
  + \frac{T n_i A_i}{\kappa n_0}
  + E^{\mathrm{Coul}}_i
  \nonumber \\ && + \frac{Z_i P^{\mathrm{Coul}}_i}{n_e}
  + Z_i \mu_e \, ,
  \label{eq:mui}
\end{eqnarray}
where $\mu_{i,\mathrm{cl}} \equiv (\partial
f_{i,\mathrm{cl}})/(\partial n_i) $.
For the nucleons
\begin{equation}
  \mu_n = \mu_{n,\mathrm{Hom}} +  \frac{T n_i}{\kappa n_0}
  \label{eq:mun}
\end{equation}
and 
\begin{equation}
  \mu_p = \mu_{p,\mathrm{Hom}} +  \frac{T n_i}{\kappa n_0} +
  \mu_e + \frac{P^{\mathrm{Coul}}}{n_e},
  \label{eq:mup}
\end{equation}
where $\mu_{x,\mathrm{Hom}} \equiv (\partial
f_{\mathrm{Hom}})/(\partial n_x) $ and $P^{\mathrm{Coul}} \equiv
\sum_i P^{\mathrm{Coul}}_i$. Note that because we have not included
electrons as separate degrees of freedom in Eq.~\eqref{eq:f_full}, the
electron chemical potential appears in Eq.\ \eqref{eq:mup}. Thus, our
chemical potentials above match those in Ref.~\cite{Hempel10}. Using
the Saha equation, we find
\begin{eqnarray}
  \mu_{i,\mathrm{cl}} &=&
  \left(\frac{A_i}{n_0}\right) f_{\mathrm{Hom}} + \mu_{n,\mathrm{Hom}}
  \left[ N_i - \left(\frac{A_i n_n}{\xi n_0 }\right) \right]
  \qquad
  \nonumber \\ && \hspace{-.2in}
  + \mu_{p,\mathrm{Hom}}
  \left[ Z_i - \left(\frac{A_i n_p}{\xi n_0 }\right) \right]
  + T \ln \kappa - E^{\mathrm{Coul}}_i,
  \label{eq:muiclx}
\end{eqnarray}
which gives us a recipe for computing the free energy for each
nucleus. Using $P_{\mathrm{Hom}} \equiv -f_{\mathrm{Hom}} +
\mu_{n,\mathrm{Hom}} n_n^{\prime}
\label{key}+ \mu_{p,\mathrm{Hom}} n_p^{\prime}$, we can rewrite this result
slightly
\begin{equation}
  \mu_{i,\mathrm{cl}} = 
  -V_i P_{\mathrm{Hom}} +
  N_i \mu_{n,\mathrm{Hom}} + Z_i \mu_{p,\mathrm{Hom}}
  + T \ln \kappa - E^{\mathrm{Coul}}_i \, .
  \label{eq:muicl}
\end{equation}
The excluded volume effect reflected in the $T \ln \kappa$ and
$-P_{\mathrm{Hom}} V_i$ terms suppresses the number density of nuclei
near saturation densities. 

At a fixed grid point in $(n_B,Y_e,T)$ space, given $n_n$ and $n_p$,
we can compute $\kappa$ and $\xi$ using their definitions above,
compute the homogeneous matter EOS and $E^{\mathrm{Coul}}_i$ and thus
use Eq.~\eqref{eq:muicl} to compute $\mu_{i,\mathrm{cl}}$. This is then
used to compute $n_i$ and then we can solve Eqs.~\eqref{eq:cons} to
obtain the correct value of $n_n$ and $n_p$. Internally, our code
defines $x_n\equiv n_n^{\prime}/n_0$ and $x_p\equiv n_p^{\prime}/n_0$
and then solves Eqs.~\eqref{eq:cons} in terms of the variables
$\log_{10} x_n$ and $\log_{10} x_p$.

The solution of Eqs.~\eqref{eq:cons} is not unique because of the
liquid-gas phase transition and the discrete nature of the nuclei in
the distribution, so we often use neighboring points as initial
guesses and choose the solution that minimizes the free energy. Our
solver automatically decreases the step size when unphysical
configurations are encountered, but occasionally it does not converge,
especially just below the nuclear saturation density. 

We approach this with a combination of techniques, all of
which are automatically applied until a solution is found: (i)
iteratively solving for neutron and proton conservation separately
using a bracketing method (ii) using a minimizer instead of a solver
and (iii) restarting the solver with random initial points near the
initial guess.

\subsection{First derivatives}

After having solved the Saha equations for $n_i(n_n,n_p)$, it is
useful to define new ``effective'' chemical potentials for the
nucleons which include the nucleons both inside and outside nuclei
\begin{equation}
  \nu_n \equiv \left( \frac{\partial f}{\partial n_n}
  \right)_{n_p,T} \quad \mathrm{and} \quad
  \nu_p \equiv \left( \frac{\partial f}{\partial n_p}
  \right)_{n_n,T}
\end{equation}
(note that these differ from Eq.~\eqref{eq:mu_defn} in that they no longer
hold $n_i$ constant) 
which gives a new thermodynamic identity
\begin{equation}
  f(n_n,n_p,T) = -P(n_n,n_p,T) + \nu_n n_n + \nu_p n_p \, .
\end{equation}
Rewriting the free energy again
\begin{equation}
  f[n_n,n_p,\{n_i(n_n,n_p,T)\},T] \, ,
\end{equation}
which implies that the effective chemical potentials can be computed
in terms of the definitions above
\begin{eqnarray}
  \nu_x &=& \mu_x + \sum_i \mu_i
  \left(\frac{\partial n_i}{\partial n_x}\right)_{n_{\hat{x}}}
\end{eqnarray}
for both $\left\{x,\hat{x}\right\}=\left\{n,p\right\}$
and $\left\{x,\hat{x}\right\}=\left\{p,n\right\}$.
Defining
\begin{equation}
  g_j \equiv \mu_j - \mu_n N_j - \mu_p Z_j 
\end{equation}
we can take advantage of the fact that all the $g_j$ are constant to
write
\begin{equation}
  \left(\frac{\partial n_i}{\partial n_x}\right)_{n_{\hat{x}},g_j}
  = 
  - \left(\frac{\partial g_i}{\partial
    n_x}\right)_{n_i,n_{\hat{x}},g_{j\neq i}}
  \left(\frac{\partial n_i}{\partial
    g_i}\right)_{n_x,n_{\hat{x}},g_{j\neq i}}.
\end{equation}
The first derivative on the RHS can be obtained directly from
Eqs.~\eqref{eq:mui}, \eqref{eq:mun}, and \eqref{eq:mup}. The second
derivative is just an element along the diagonal of the inverse of the
matrix
\begin{equation}
  M_{ij} \equiv \left(\frac{\partial g_i}{\partial
    n_j}\right)_{n_x,n_{\hat{x}},n_{k\neq j}} \, .
\end{equation}
The numerical errors associated with inverting this large matrix
decreases the benefit of the analytical formalism. Thus we compute
$\nu_n$ and $\nu_p$ numerically for now. The entropy is easier
to compute
\begin{equation}
  s=-\left(\frac{\partial f}{\partial T}\right)_{n_n,n_p,\{n_i\}}=
  s_e+\sum_{i} s_i + \sum_{i}
  n_i \ln \kappa + \xi s_{\mathrm{Hom}}
\end{equation}
where $s_e = \partial f_e/\partial T$ and
\begin{equation}
  s_i=n_i \left(\ln \frac{\Omega_i}{n_i \lambda_i^3}\right) +
  \frac{5}{2}+ \frac{T}{\Omega_i}  \frac{d \Omega_i}{d T} \, .
  \label{eq:snuc}
\end{equation}

\subsection{Nuclei}

We use the nuclear masses from experiment~\cite{Audi12} wherever they
are available. The atomic mass tables usually include an empirical
bounded electron contribution term $a_{el} Z^{2.39}$, which is
subtracted before the binding energy is calculated. We use the
theoretical masses from Ref.~\cite{Moller95} for nuclei which do not
have experimental mass measurements up to the neutron and proton drip
lines.
We use the experimental or theoretical spins tabulated in
Ref.~\cite{Goriely07fe}.
Finally, we limit
$Z<7N$ and $N<7Z$ in order to avoid extreme nuclei which our model
likely does not describe well.

\subsection{Partition function}

The partition function we use for light nuclei and the representative heavy
nucleus follows from Ref.~\cite{Shen10}. The nuclear partition
function can be expressed as a sum of discrete states and an integral
of the level density
\begin{equation}
  \Omega_i=(2J+1) + \int_{E_d}^{E_t} \rho(E) \exp(-E/T) \, ,
\end{equation}
where the level density $\rho(E)$ is the backshifted Fermi-gas formula
given below. The limits on the integral in the partition function are
determined from
\begin{eqnarray}
  &&E_d=\frac{1}{2} \mathrm{min}(S_n, S_p) \quad \mathrm{and} \\
  &&E_t=\mathrm{min}(S_n+E_R,S_p+E_R+\frac{1}{2}E_c) \, ,
\end{eqnarray}
where $S_n$ and $S_p$ are the neutron and proton separation energies.
The quantity $E_R\equiv 1/(2 M_i R^2)$
is the zero-point energy and with the nuclear radius
approximated by $R=1.25~\mathrm{fm}~(A-1)^{1/3}$. The Coulomb barrier
is $E_c\equiv (Z-1) \alpha /R$. When either $S_n $ or $S_p$ is
negative, the contribution of the level density to the partition
function is neglected.

The expression for the level density begins by defining a
backshift parameter $\delta$ for each nucleus. The prescription from
Ref.~\cite{Fowler78np} is
\begin{eqnarray}
  &&Z \leqslant 30: \delta=\delta_p-80/A \\
  &&Z \geqslant 30: \delta=\delta_p-80/A-0.5 
\end{eqnarray}
with $\delta_p=(11 A^{-1/2}~\mathrm{MeV}) [1+(1/2)(-1)^Z +1/2
  (-1)^N]$. We will also need the level density parameter, $a$, for
which an approximate model is
\begin{eqnarray}
  &&Z \leqslant 30: a=0.052~\mathrm{MeV}^{-1}~A^{1.2} \\
  &&Z \geqslant 30: a=0.125~\mathrm{MeV}^{-1}~A \, .
\end{eqnarray}
Finally, different expressions are used for the level density
depending on the relative size of $\delta$ and $E_d$. When $\delta$ is
smaller than $E_d$, the level density has the expression
\begin{equation}
  \rho (E) = \frac{\pi}{12} \frac{\exp(2 \sqrt{a U})}{a^{1/4} U^{5/4}},
  \label{eq:ld1}
\end{equation}
where $U=E-\delta$.

When $\delta$ is larger than $E_d$, $\delta $ is set to $E_d$ (so that
$U=E-E_d$) and the level density is
\begin{equation}
  \rho (E) =
  C \exp(U/T_c),
\label{eq:ld2}
\end{equation}
where
\begin{eqnarray}
  &&\frac{1}{T_c} = \frac{5}{4} \frac{1}{\delta} +
  \frac{\sqrt{a}}{\sqrt{\delta}} \, , \,\mathrm{and} \\ \nonumber
  && C= \frac{\sqrt{\pi}}{12} a^{-1/4} {\delta}^{-5/4}
  \exp \left(\frac{5}{4} + \sqrt{a \delta}\right) \, .
\end{eqnarray}
The derivative of the partition function with respect to the
temperature is required for computing the entropy (see Eq.~\eqref{eq:snuc}),
and this is straightforward to compute analytically.

\section{Results}

While our EOS formalism is designed to be used for any physical values
of the parameters, we precompute 7 tables and present results based on
those parameterizations. The parameters $\{i_{\mathrm NS},
i_{\mathrm{Skyrme}}, \alpha, a, L~(\mathrm{MeV}), S~(\mathrm{MeV}),
\phi \}$ are (i) the choice of high-density EOS parameterization
selected from a discrete set of Markov-chain samples constructed in
Ref.\ \cite{Steiner15un}, (ii) the choice of Skyrme effective
interaction selected from 1000 samples generated from the posterior
probability distribution in Ref.\ \cite{Kortelainen14}, (iii and iv)
the power and prefactor in Eq.\ \eqref{qmcnm} for the neutron matter
equation of state, (v and vi) the symmetry energy slope parameter, the
symmetry energy, and (vii) the speed of sound at the largest density
we consider, $n_B = 2~\mathrm{fm}^{-3}$. The parameters of the seven
equation of state tables are listed in Table~\ref{tab:params}. The
fiducial EOS is consistent with the most probable neutron star mass
and radius while having moderate $S$ and $L$ (see details in
Ref.~\cite{Du19hd}). The Skyrme parameters for our fiducial EOS are
listed in Table~\ref{tab:skyrme}. In the following, the figures are
demonstrated for our fiducial EOS.

\begin{table}
  \centering
  \begin{tabular}{c|ccccccc}
    & $i_{\mathrm{NS}}$ & $i_{\mathrm{Skyrme}}$ &  $\alpha$ & $a$ &
    $L~(\mathrm{MeV})$ & $S~(\mathrm{MeV})$ & $\phi$ \\
    \hline
    ${\mathrm{fiducial}}$ &
    470 & 738 & 0.5 & 13.0 & 62.4 & 32.8 & 0.9 \\
    ${\mathrm{large}\,M_{\mathrm{max}}}$ &
    783 & 738 & 0.5 & 13.0 & 62.4 & 32.8 & 0.9 \\
    ${\mathrm{small}\,R}$ &
    214 & 738 & 0.5 & 13.0 & 62.4 & 32.8 & 0.9 \\
    ${\mathrm{smaller}\,R}$ &
    256 & 738 & 0.5 & 13.0 & 62.4 & 32.8 & 0.9 \\
    ${\mathrm{large}\,R}$ &
    0 & 738 & 0.5 & 13.0 & 62.4 & 32.8 & 0.9 \\
    ${\mathrm{small}\,SL}$ &
    470 & 738 & 0.5 & 13.0 & 23.7 & 29.5 & 0.9 \\
    ${\mathrm{large}\,SL}$ &
    470 & 738 & 0.5  & 13.0 & 100.0 & 36.0 & 0.9 \\
    \hline
  \end{tabular}
  \caption{Parameters for the EOS tables generated
    for this work.\label{tab:params}.}
\end{table}

\begin{table}
  \centering
  \begin{tabular}{c|c}
    $t_0$ & 
    $-$2719.7~$\mathrm{MeV}~\mathrm{fm}^3$ \\
    $t_1$ & 
    417.64~$\mathrm{MeV}~\mathrm{fm}^5$ \\
    $t_2$ & 
    $-$66.687~$\mathrm{MeV}~\mathrm{fm}^5$ \\
    $t_3$ & 
    15042~$\mathrm{MeV}~\mathrm{fm}^{3(1+\epsilon)}$ \\
    $x_0$ &
    0.16154 \\
    $x_1$ & 
    $-$0.047986 \\
    $x_2$ & 
    0.027170 \\
    $x_3$ & 
    0.13611 \\
    $\epsilon$ &
    0.14416 \\
  \end{tabular}
  \caption{Parameters for Skyrme Hamiltonian $i_{\mathrm{Skyrme}}=738$
    \label{tab:skyrme}}
\end{table}

\subsection{Composition of Hot and Dense Matter}

\begin{figure}
  \includegraphics[width=3.1in]{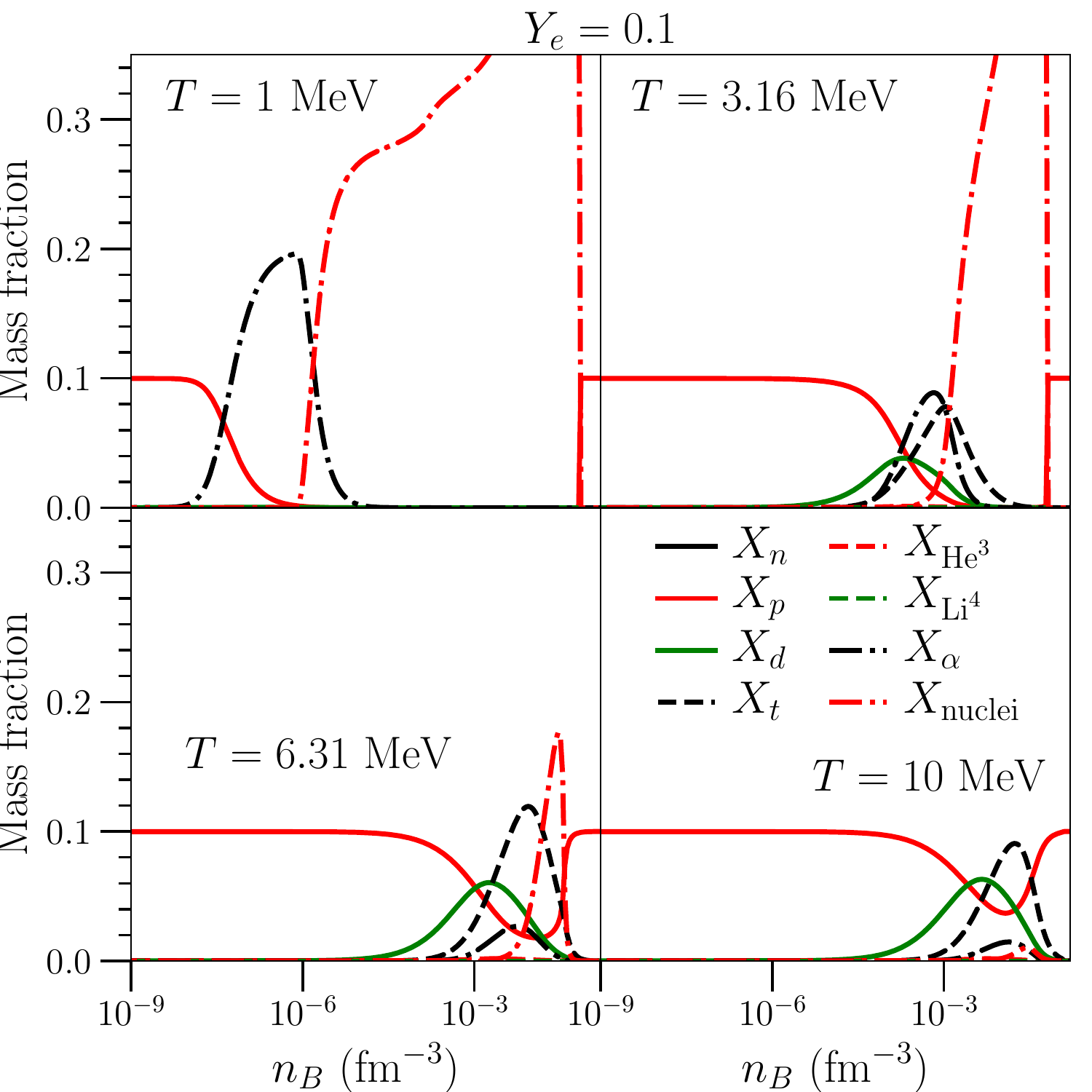}
  \includegraphics[width=3.1in]{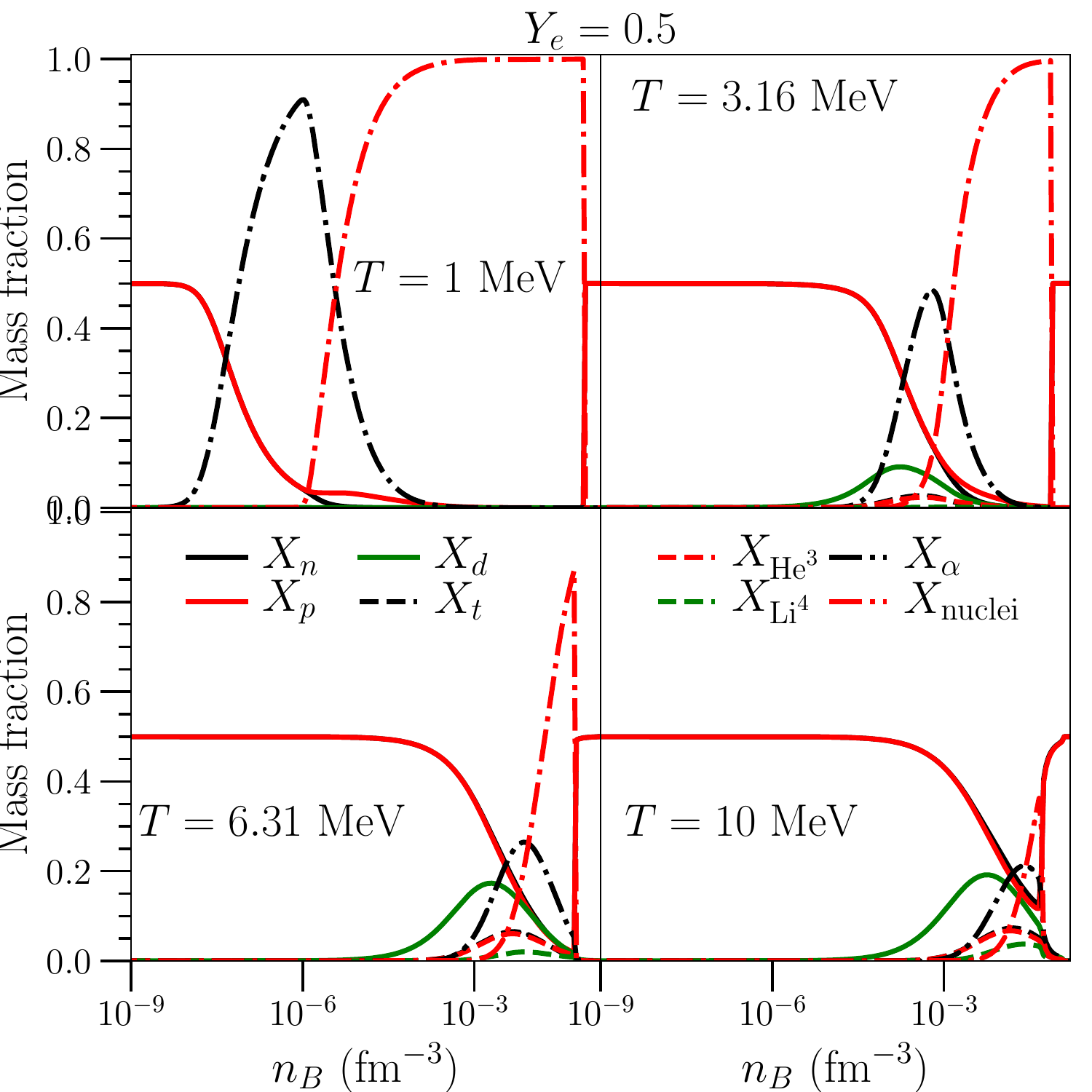}
  \caption{Baryon number fractions $X_i$ for protons, light nuclei,
    and a sum over heavy nuclei, as a function of density for
    $Y_e=0.1$ and $Y_e=0.5$ for four temperatures. In the top four
    panels, the neutron baryon number fraction is omitted to help make
    the heavy nuclei more visible. In the bottom four panels, the
    neutron mass fraction is hidden behind the proton mass fraction at
    low densities where these two quantities coincide. The right edge
    of the plots is chosen to be $n_B=n_0$ and nuclei always disappear
    at a baryon density below $n_0$ (independent of electron fraction
    or temperature).
  \label{fig:lightnuc}}
\end{figure}

In Fig.~\ref{fig:lightnuc}, the baryon number fraction of free neutrons,
protons, light nuclei and heavy nuclei are plotted as a function of
baryon density for $Y_e=0.1$ and $Y_e=0.5$. The baryon number
fraction of species $i$ is
\begin{equation}
  X_i \equiv n_i A_i / n_B,
\end{equation}
where $n_i$ is the number per unit volume for species $i$ and $A_i$ is
the number of baryons in species $i$. Eq.~\eqref{eq:cons} ensures
$\sum_i X_i = 1$. The quantity $X_{\mathrm{nuclei}}$ is defined by
\begin{equation}
  X_{\mathrm{nuclei}}\equiv
  1-X_n-X_p-X_d-X_t-X_{\alpha}-X_{^{3}\mathrm{He}}-X_{^{4}\mathrm{Li}}.
\end{equation}
At low densities, the system consists of only protons and neutrons.
For $Y_e=0.1$ and $T=1.0~\mathrm{MeV}$, as density increases, the mass
fraction of alpha partices rises to around 0.2 for $n_B$ between
$10^{-7}$ and $10^{-6}~\mathrm{fm}^{-3}$. Above
$10^{-6}~\mathrm{fm}^{-3}$, the light nuclei are gradually replaced by
heavy nuclei. The transition density from light to heavy nuclei
increases as temperature increases. For $Y_e=0.5$, alpha particles are
even more prominent at lower densities and heavier nuclei dominate
more strongly near the transition to nucleonic matter. For higher
temperature (but independent of electron fraction), the region of
light and heavy nuclei gradually merge to a single peak.

\begin{figure}
  \includegraphics[width=3.1in]{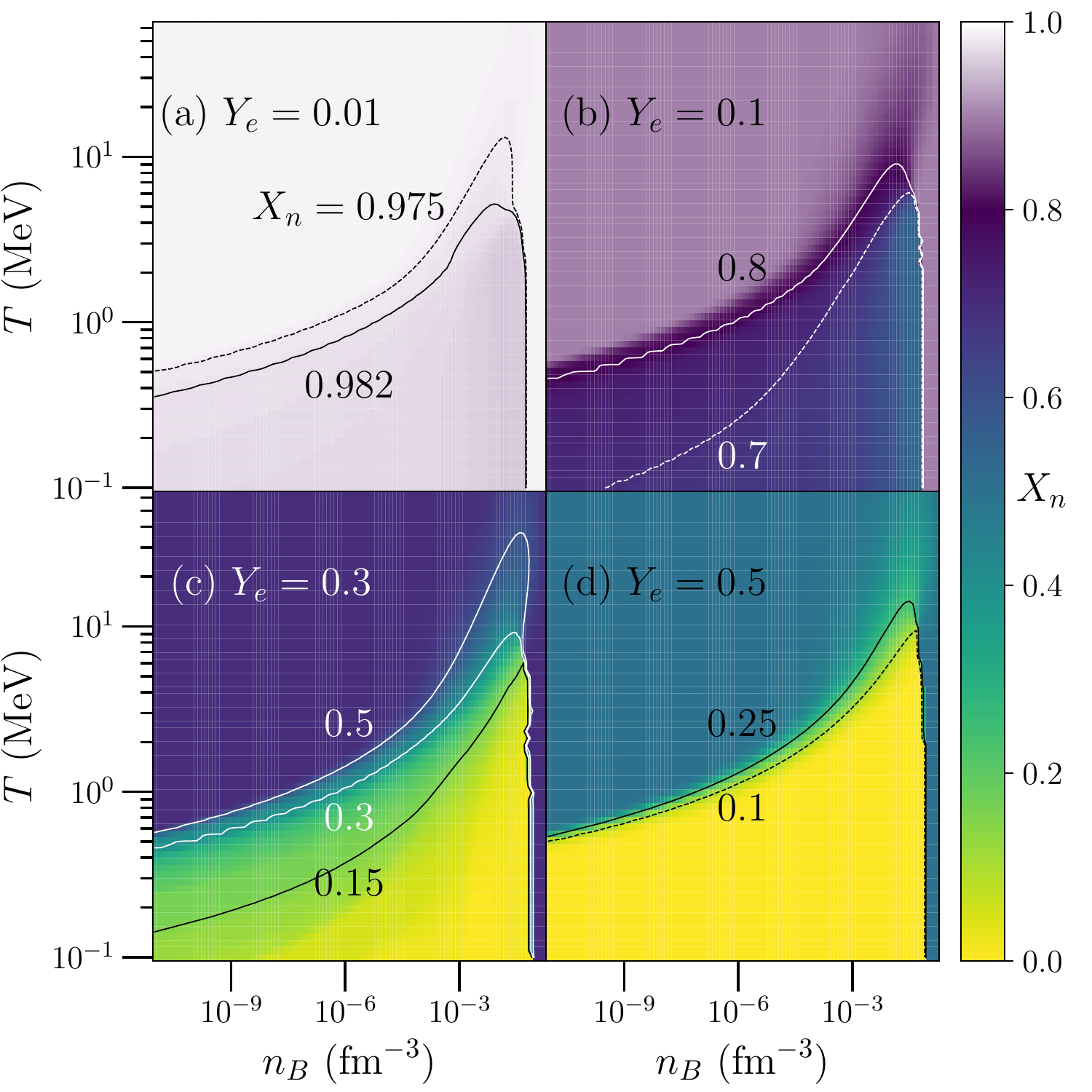}
  \includegraphics[width=3.1in]{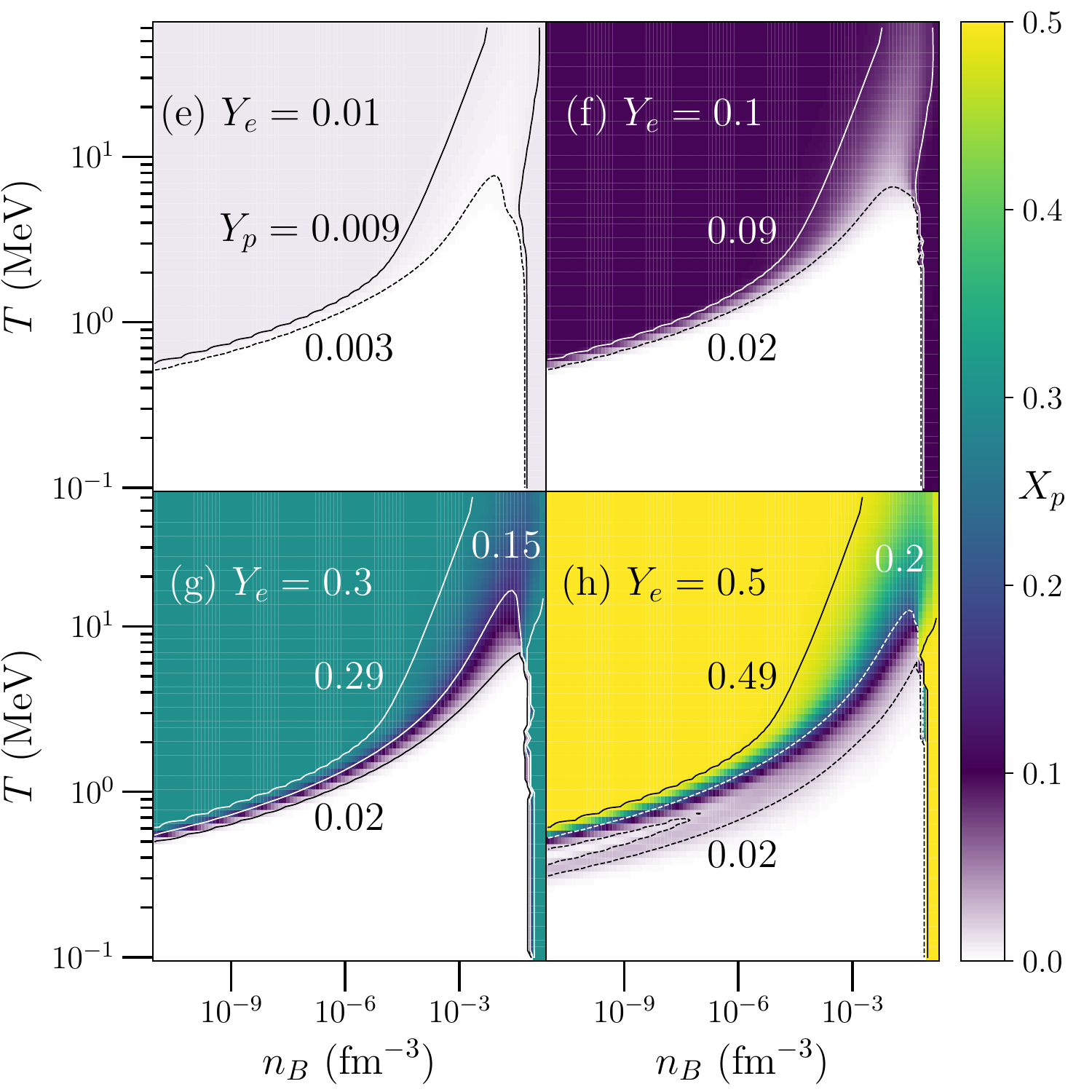}
  \caption{Baryon number fractions $X_n$ and $X_p$ as a function of
    baryon density and temperature for $Y_e=0.01$, 0.1, 0.3, 0.5,
    respectively. The right edge of the plots is
    chosen to be $n_B=n_0$ where nuclei disappear (independent of
    electron fraction or temperature). 
  \label{fig:xnp}}
\end{figure}

\begin{figure}[h]
  \includegraphics[width=3.1in]{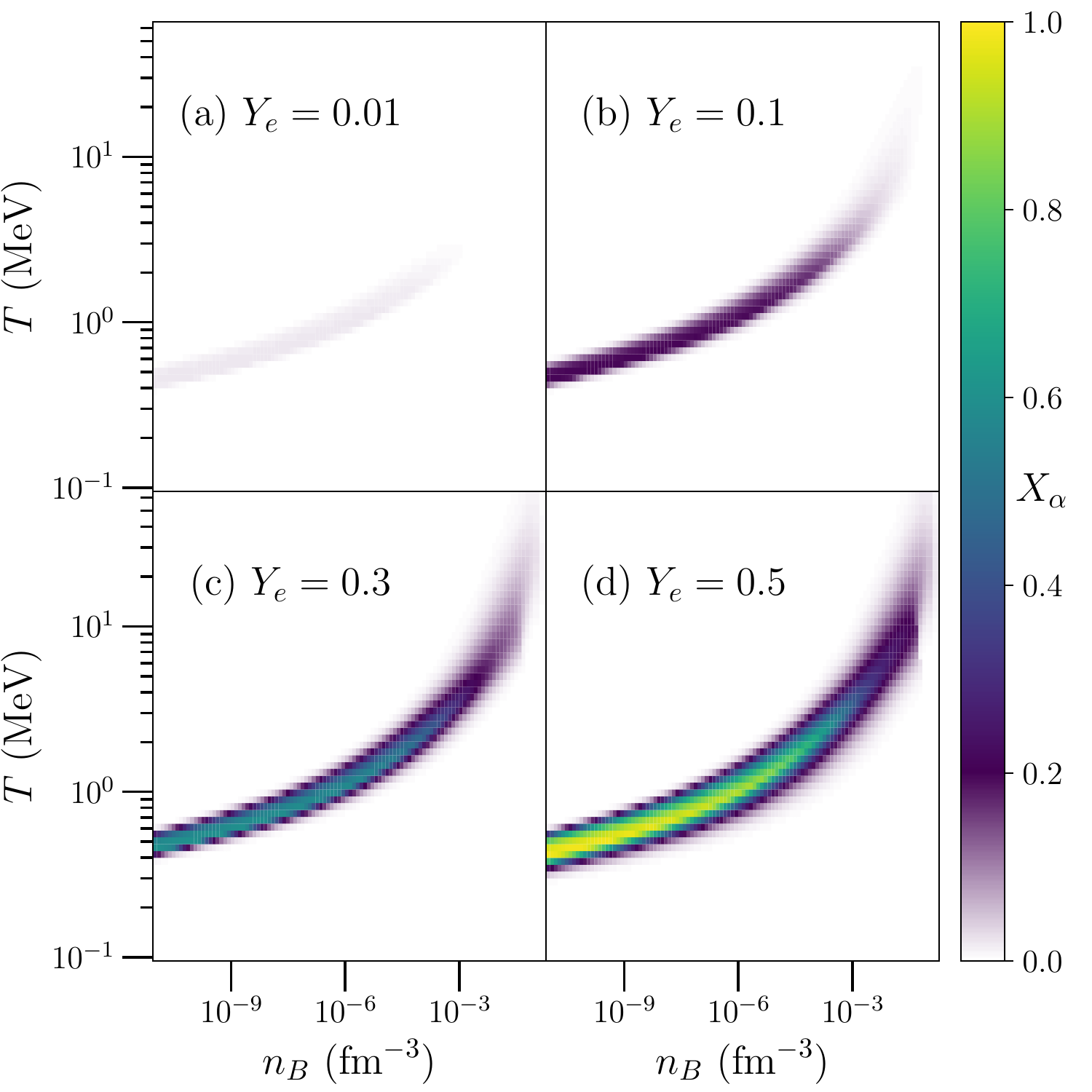}
  \includegraphics[width=3.1in]{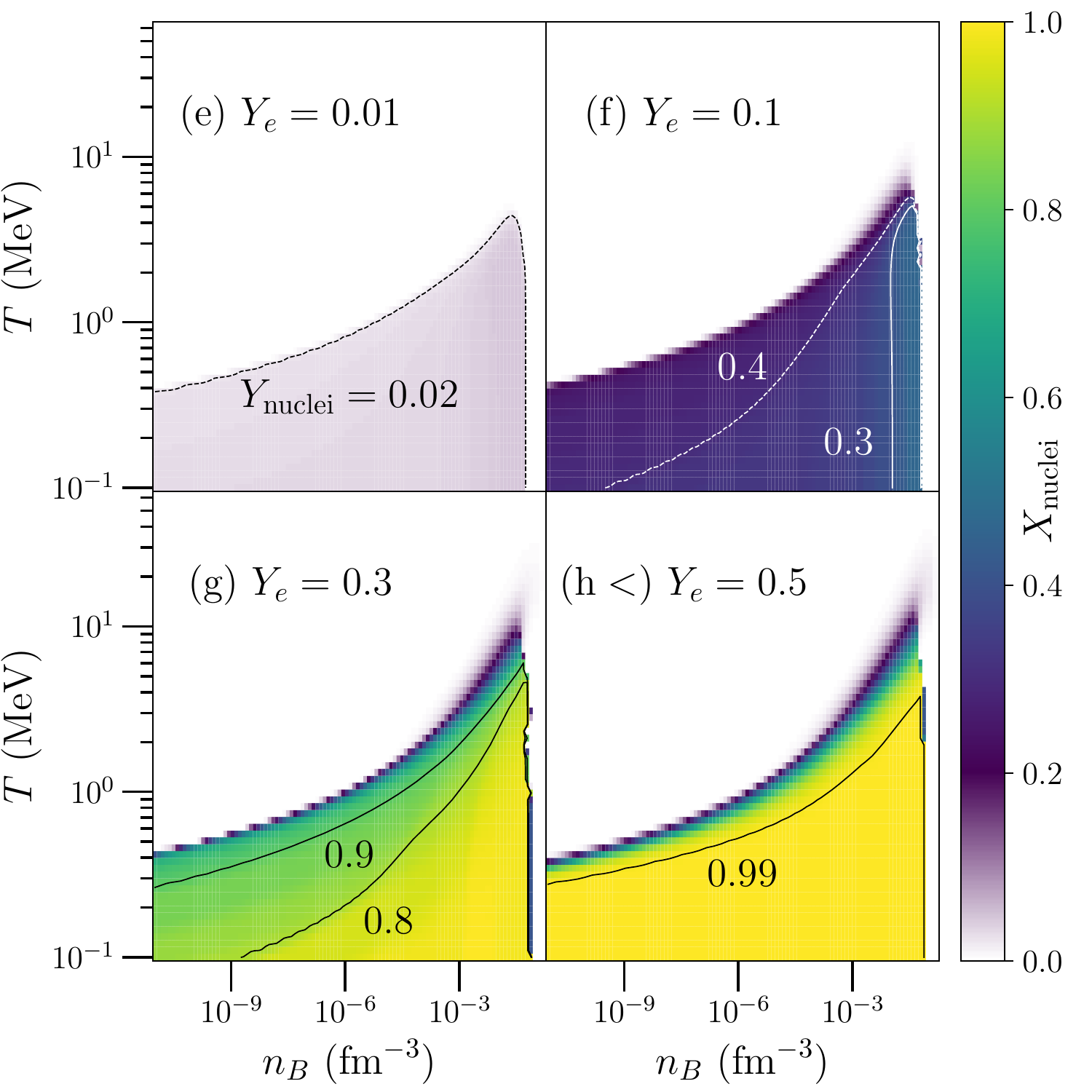}
  \caption{Baryon number fractions $X_{\alpha}$ and
    $X_{\mathrm{nuclei}}$ as a function of baryon
    density and temperature for $Y_e=0.01$, 0.1, 0.3, 0.5,
    respectively. The right edge of the plots is chosen to be
    $n_B=n_0$ where nuclei disappear (independent of electron fraction
    or temperature).
  \label{fig:xa}}
\end{figure}

Fig.~\ref{fig:xnp} shows baryon number fractions $X_n$, $X_p$ and
Fig.~\ref{fig:xa} shows baryon number fractions $X_{\alpha}$,
$X_{\mathrm{nuclei}}$ as a function of baryon density and temperature.
Near $Y_e=0.5$ and at low temperatures, the system consists almost
entirely of heavy nuclei. As the temperature increases, the
non-uniform clusters transform to uniform matter. On the other hand,
as $Y_e$ decreases, nuclei are replaced by free neutrons. The critical
temperature of the gas-liquid phase transition is around several to
tens of MeV depending on the proton fraction.

\begin{figure}[h]
  \includegraphics[width=3.1in]{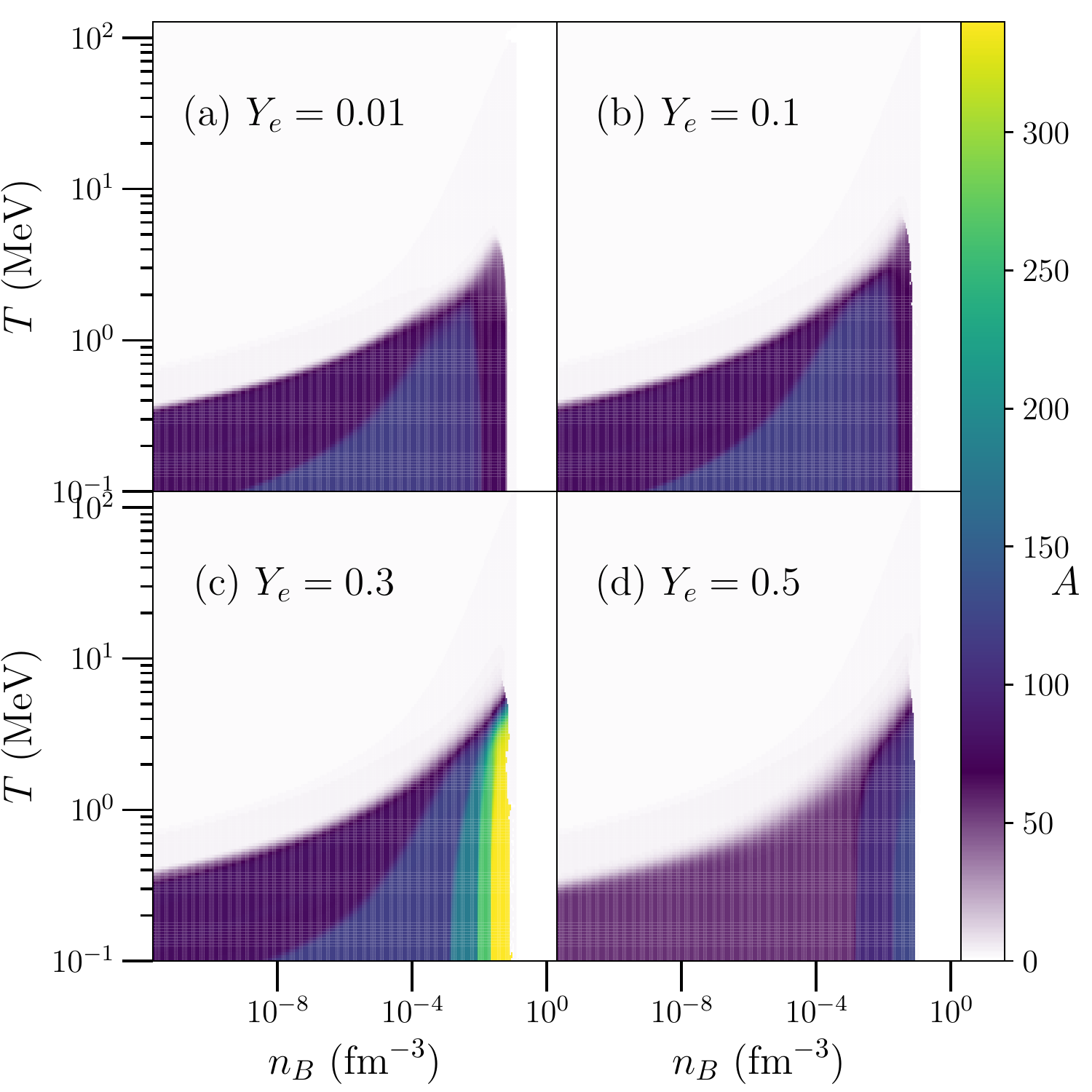}
  \includegraphics[width=3.1in]{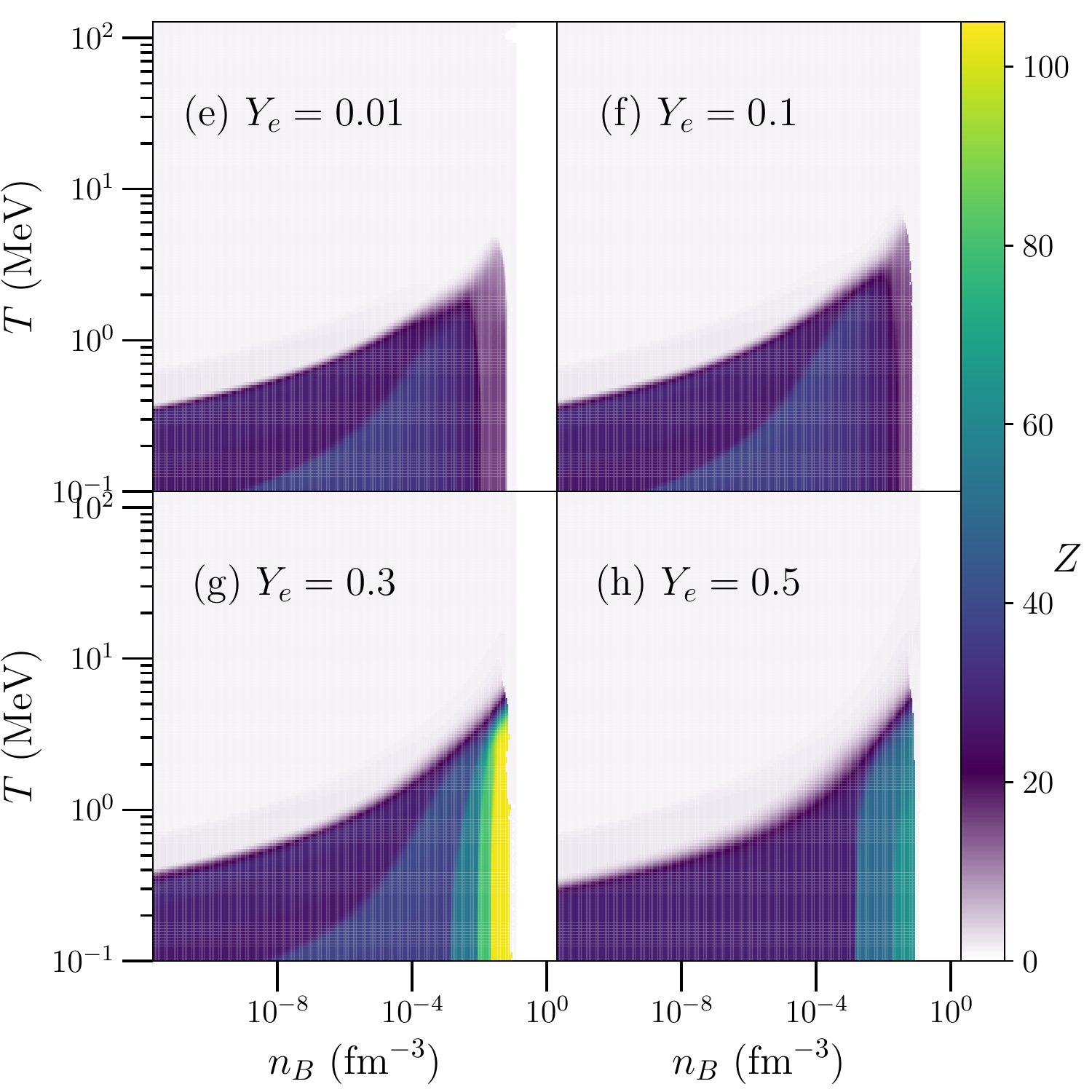}
  \caption{Average mass(top)/proton(bottom) number for
    Ye=0.01, 0.1, 0.3, 0.5, respectively.
  \label{fig:avgazye}}
\end{figure}

\begin{figure}[h]
  \includegraphics[width=3.1in]{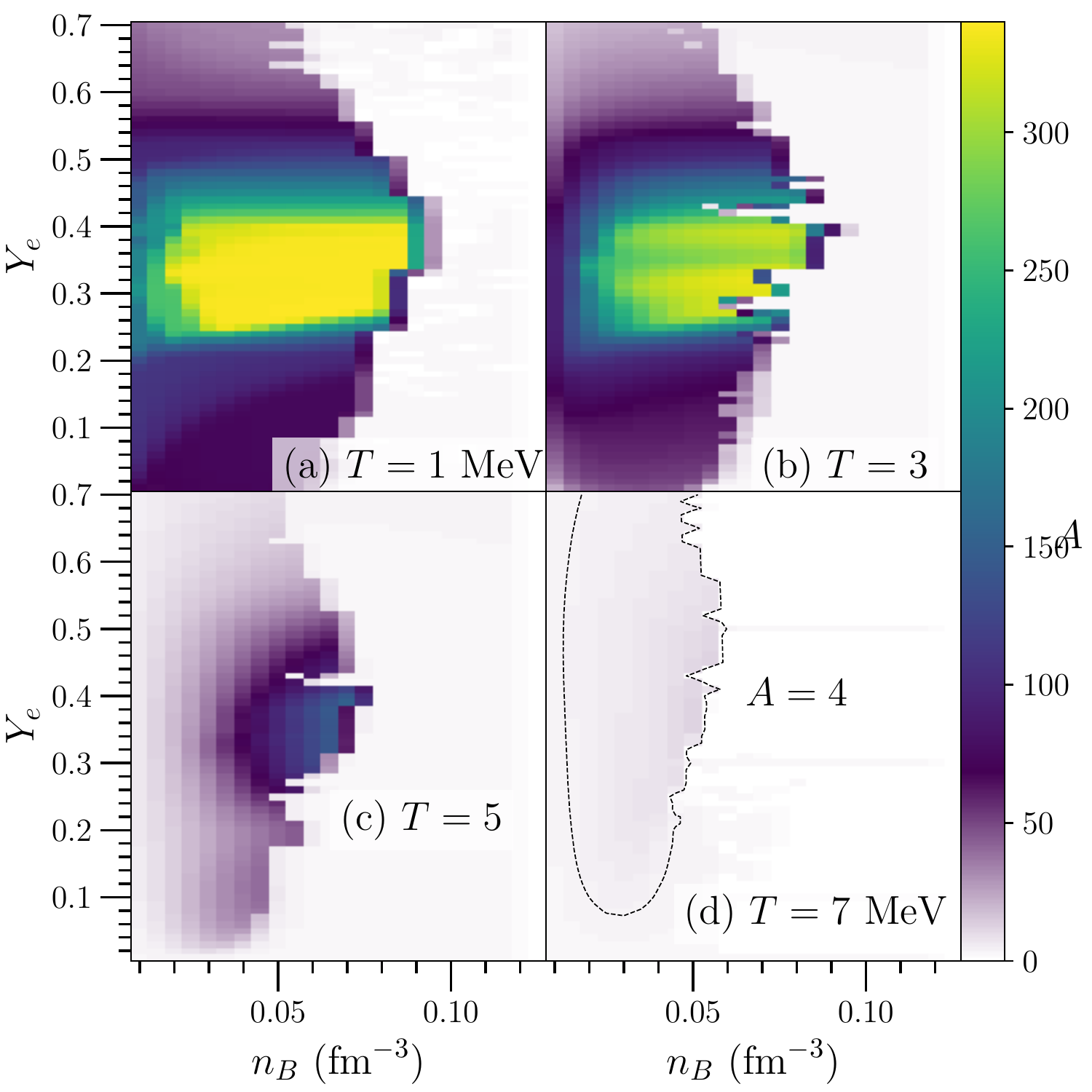}
  \includegraphics[width=3.1in]{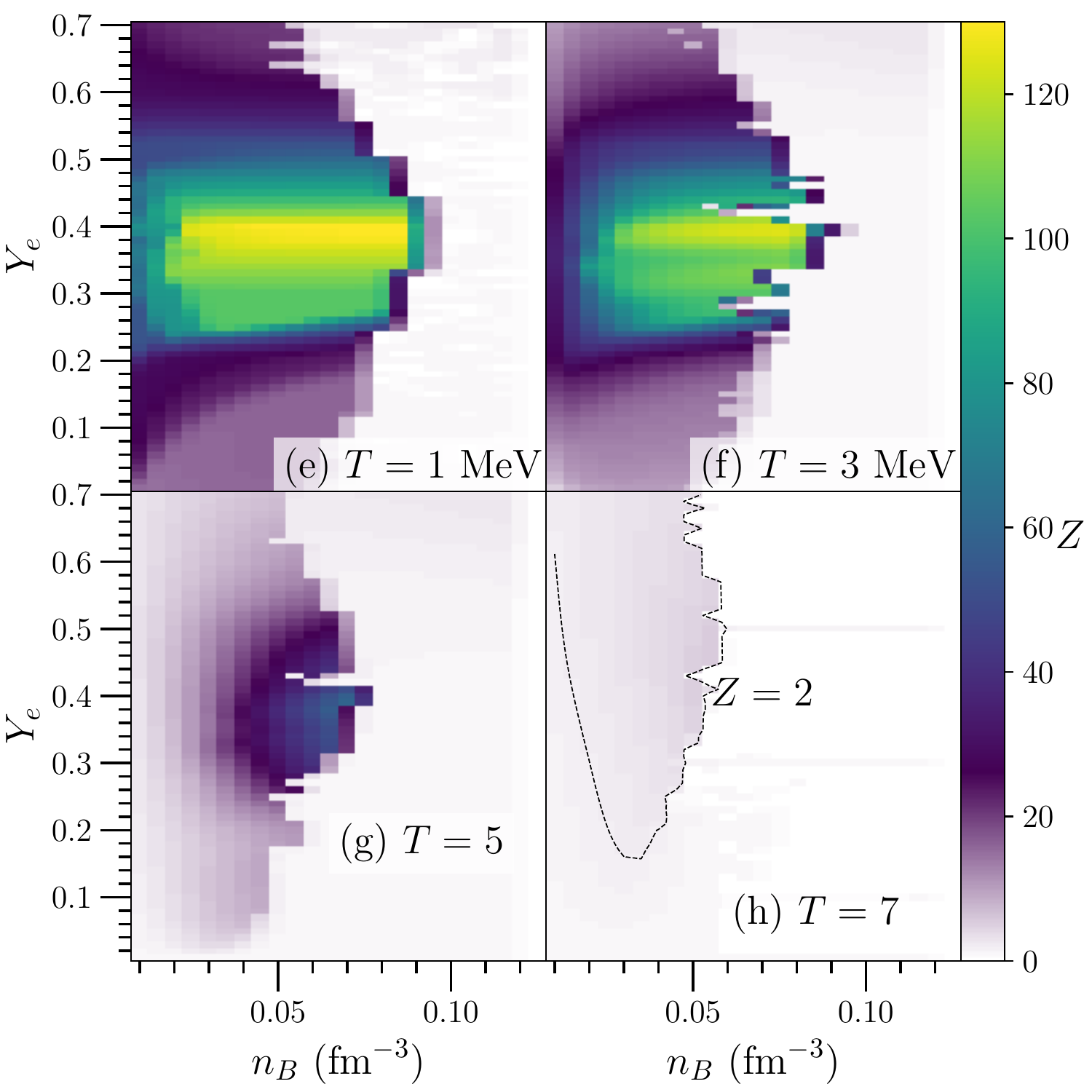}
  \caption{Average mass (top four panels) and proton (bottom four
    panels) number for T=1, 3, 5, $7~\mathrm{MeV}$, respectively.
  \label{fig:avgazt}}
\end{figure}

To compute the average proton and neutron number of nuclei,
we define
\begin{equation}
  \bar{Z} = \left( \sum_i Z_i n_i \right) \left( \sum_i n_i \right)^{-1},
\end{equation}
where this sum includes the light nuclei $d$, $t$, $\alpha$,
$^{3}\mathrm{He}$ and $^{4}\mathrm{Li}$. We define a similar quantity
$\bar{N}$, and the average nuclear mass number is then $\bar{A} \equiv
\bar{N}+\bar{Z}$. Fig.~\ref{fig:avgazye} shows $\bar{A}$ and $\bar{Z}$
as a function of baryon density and temperature. The maximum A for our
EOS is limited to about 340. For symmetric nuclear matter, $\bar{A}$
reaches the upper limit we set. For smaller electron fractions, the
maximum mass number decreases to 120 as neutrons leave nuclei to form
a gas. The shell structure of nuclei is evident in the figures as
rapid color changes. As baryon density increases, $\bar{A}$ rises to
several plateaus. Fig.~\ref{fig:avgazt} shows the charge and mass
number of nuclei as a function of density and electron fraction at
four fixed temperatures. The transition density from inhomogeneous
matter to homogeneous matter is not independent of proton fraction, as
observed in microscopic calculations of the equation of state
\cite{Fiorilla:2011sr,Wellenhofer15to}. The transition density is
largest near $Y_e \approx 0.4$, which is to be expected since heavy
laboratory nuclei have a similar proton fraction. At higher
temperatures nuclei disappear as we approach the liquid gas
transition.

\subsection{Comparison with other EOSs}
\begin{figure*}
  \includegraphics[width=3.1in]{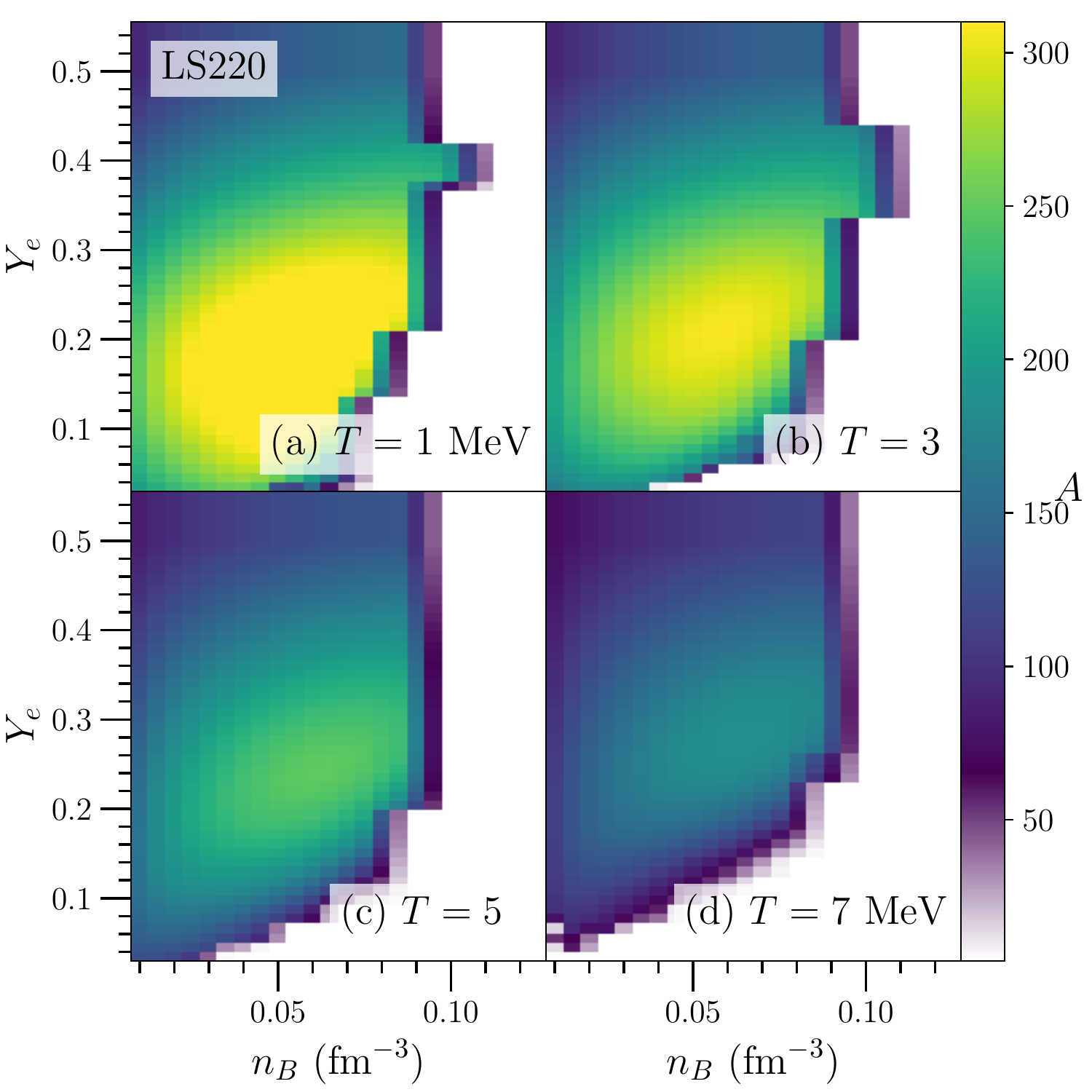}
  \includegraphics[width=3.1in]{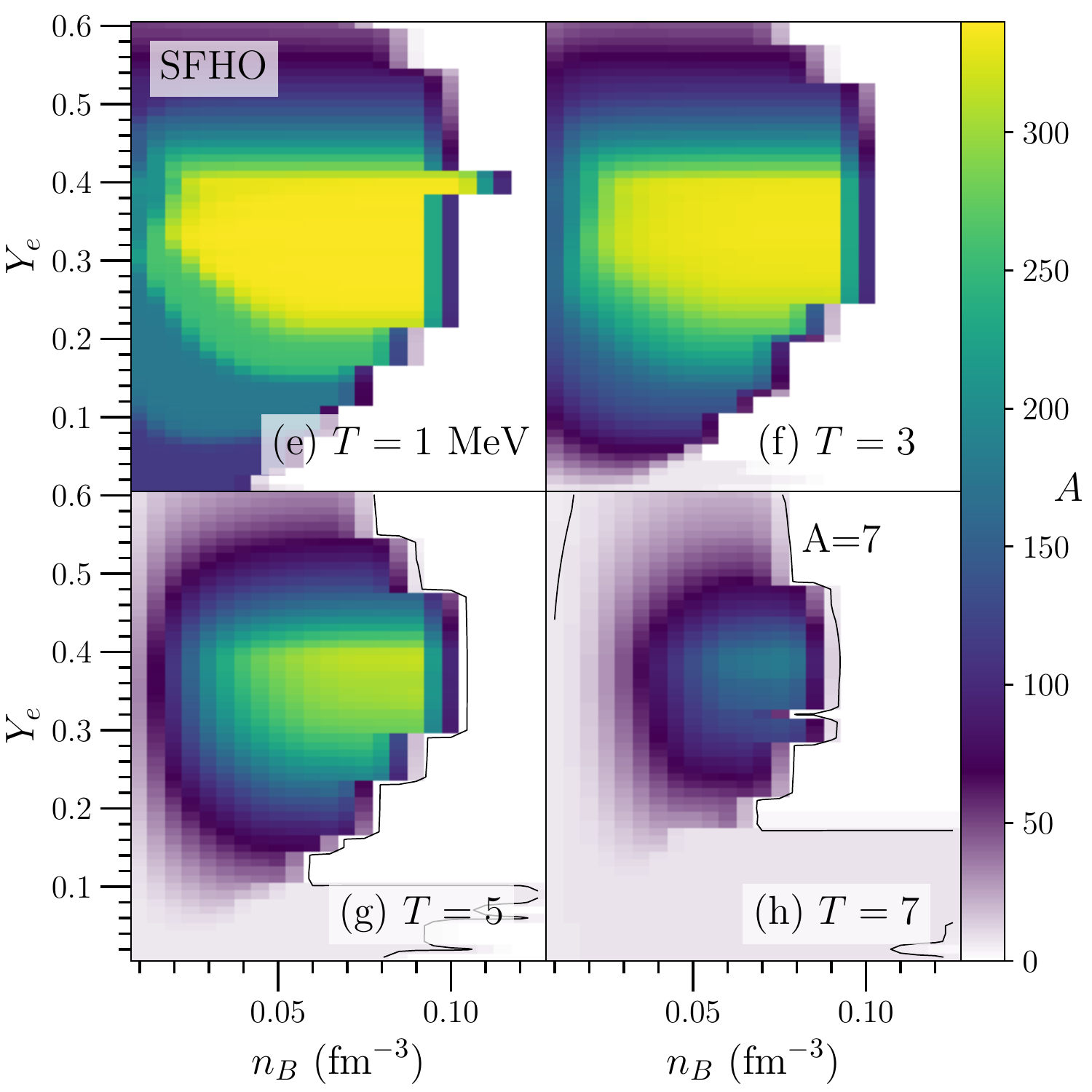}
  \includegraphics[width=3.1in]{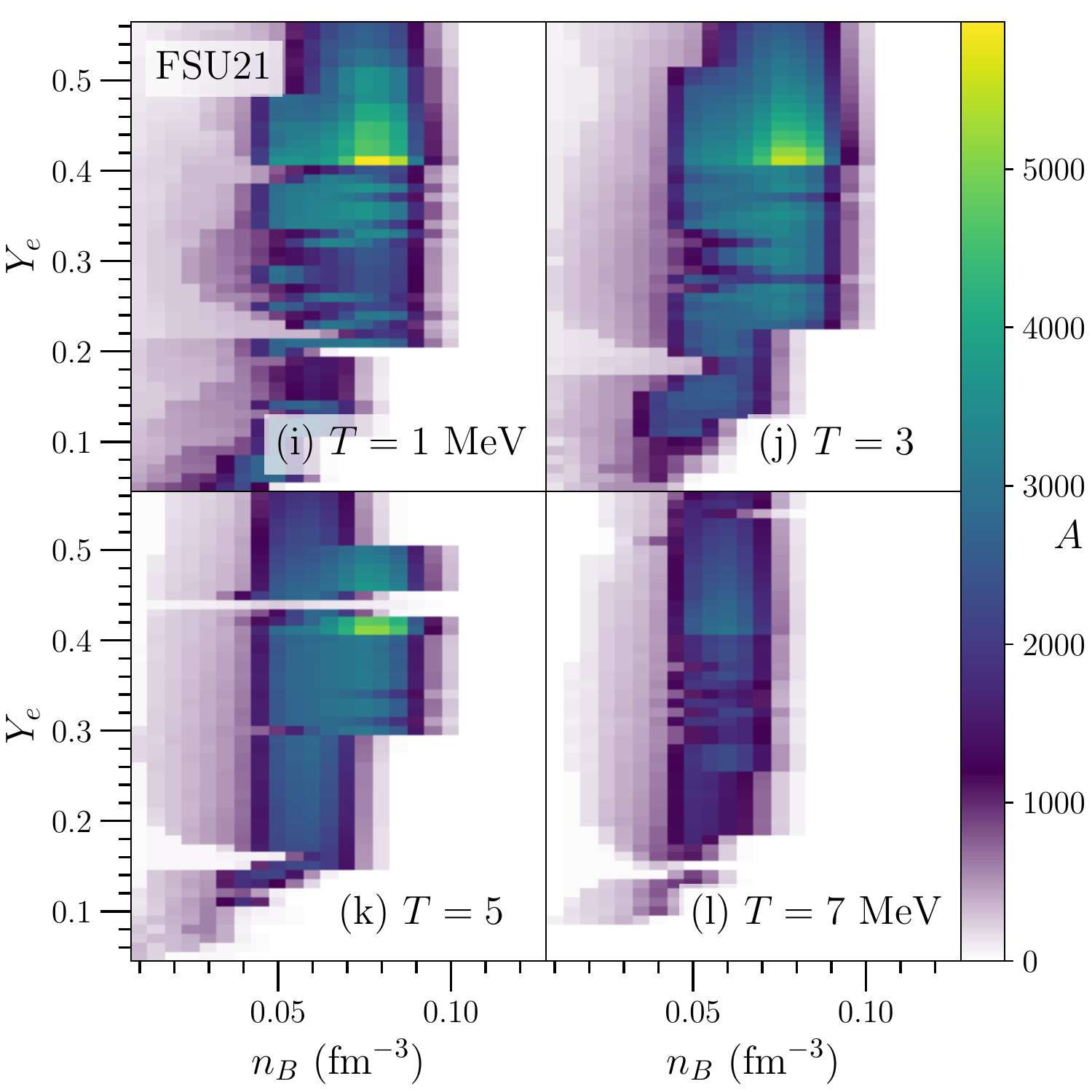}
  \includegraphics[width=3.1in]{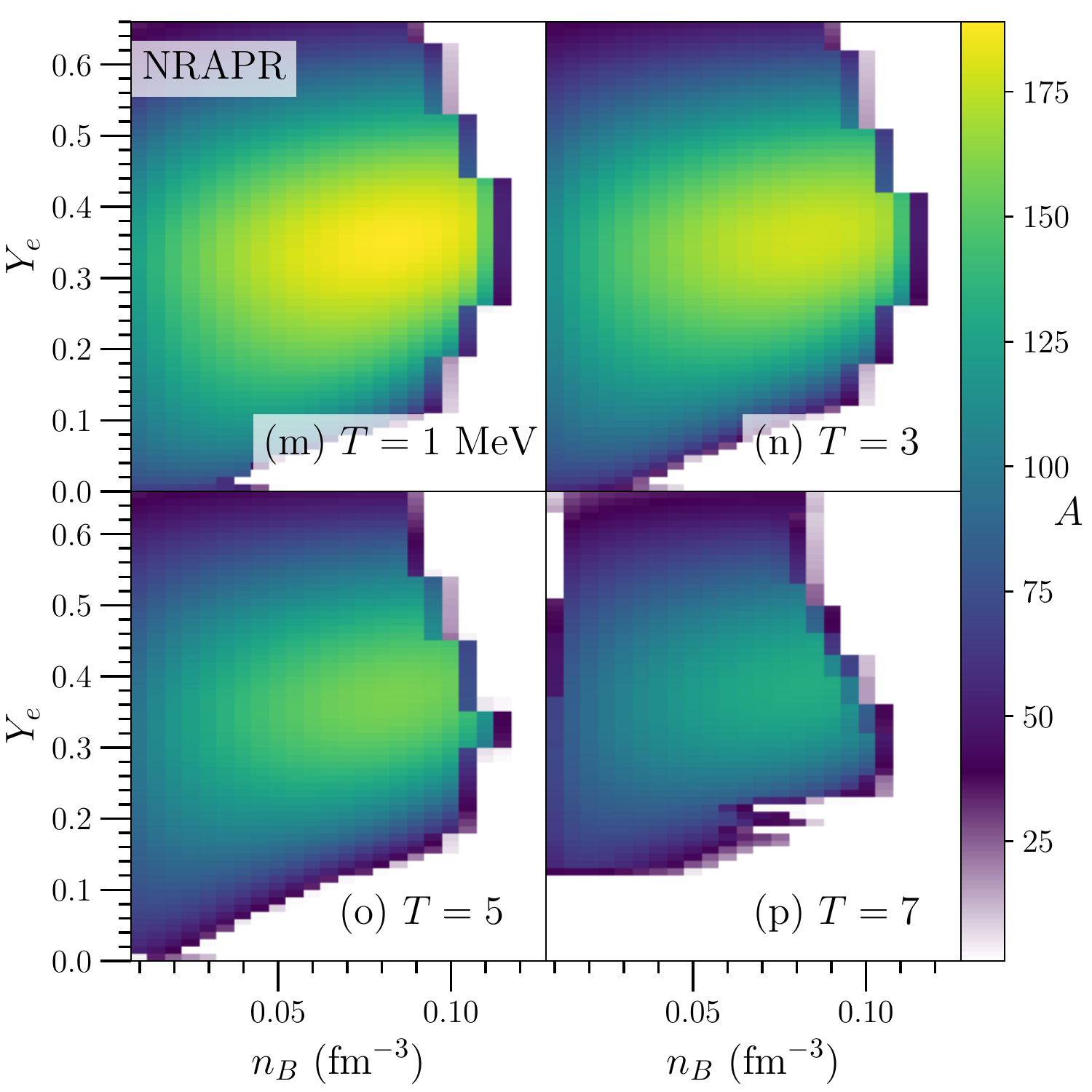}
  \includegraphics[width=3.1in]{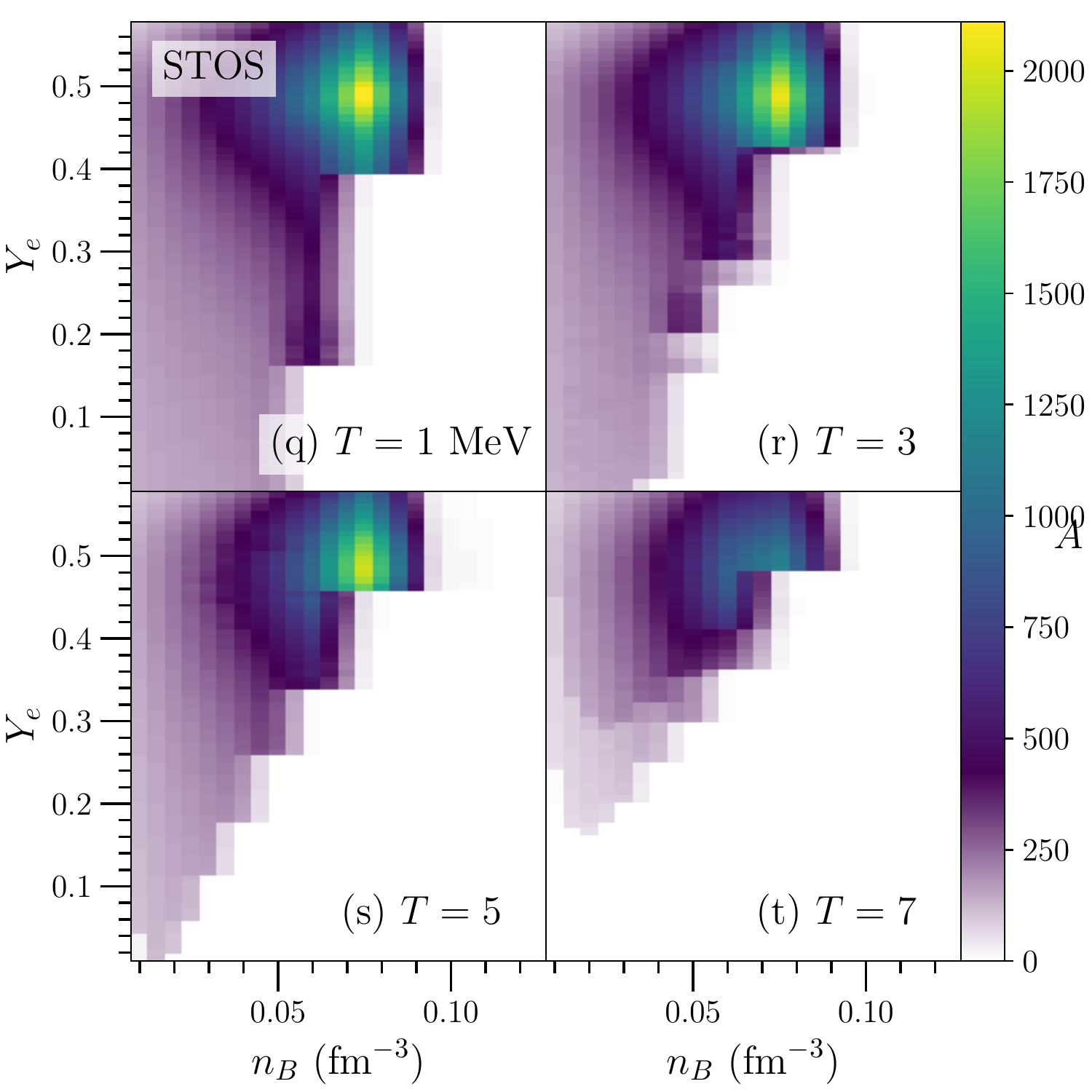}
  \includegraphics[width=3.1in]{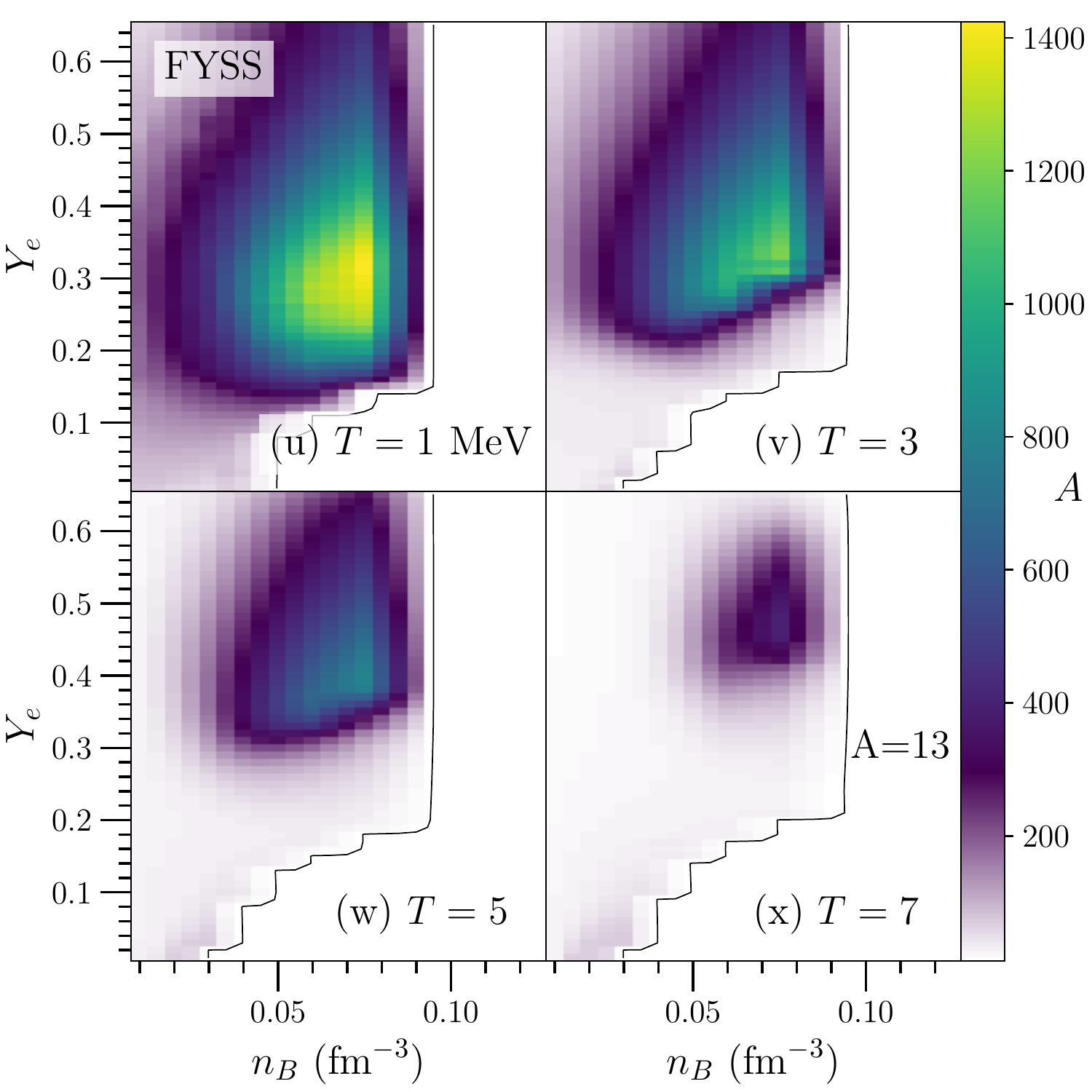}
  \caption{Average mass number for T=1, 3, 5, and $7~\mathrm{MeV}$,
    for the LS220, SFHO, FSU21, NRAPR, STOS, and FYSS EOSs.
  \label{fig:othereos}}
\end{figure*}

Fig.~\ref{fig:othereos} shows the average mass number $A$ as a
function of baryon density and temperature for several other EOSs:
LS220~\cite{Lattimer91}, SFHO~\cite{Steiner13cs}, FSU21~\cite{Shen11},
NRAPR~\cite{Schneider19es}, STOS~\cite{Shen98}, and
FYSS~\cite{Furusawa_2011}. Note that these results were interpolated
from the files created by Ref.~\cite{OConnor10}
(and stored at {\tt{stellarcollapse.org}}), and thus details
may differ slightly from the original files. Significant differences
can be found among these plots for the predictions of mass number in
inhomogeneous phase. The plots fall into two categories. STOS, FSU21
and FYSS allow nuclei with maximum mass number around several
thousand, while LS220, NRAPR and SFHo limit A below several hundred.
There is also some variation between models in the $Y_e$ dependence of
the phase transition between nuclei and nuclear matter. In FSU21 and
FYSS, the phase transition is nearly $Y_e$-independent. Note that
different panels have different maximum values of $Y_e$, and this
impacts the apparent shape of the transition to nucleonic matter. The
STOS, FSU21, and FYSS tables all include a pasta phase before
transitioning to homogeneous matter, and this also complicates the
comparison. The inclusion of the pasta phase, in general, decreases binding
energy and therefore favors a late transition to homogeneous matter. Note
however that the difference of the mass number between EOS tables does
not strongly impact the thermodynamic quantities such as the pressure
and entropy~\cite{Burrows84}.

\subsection{Nuclear distribution}
\begin{figure}[h]
  \includegraphics[width=3.1in]{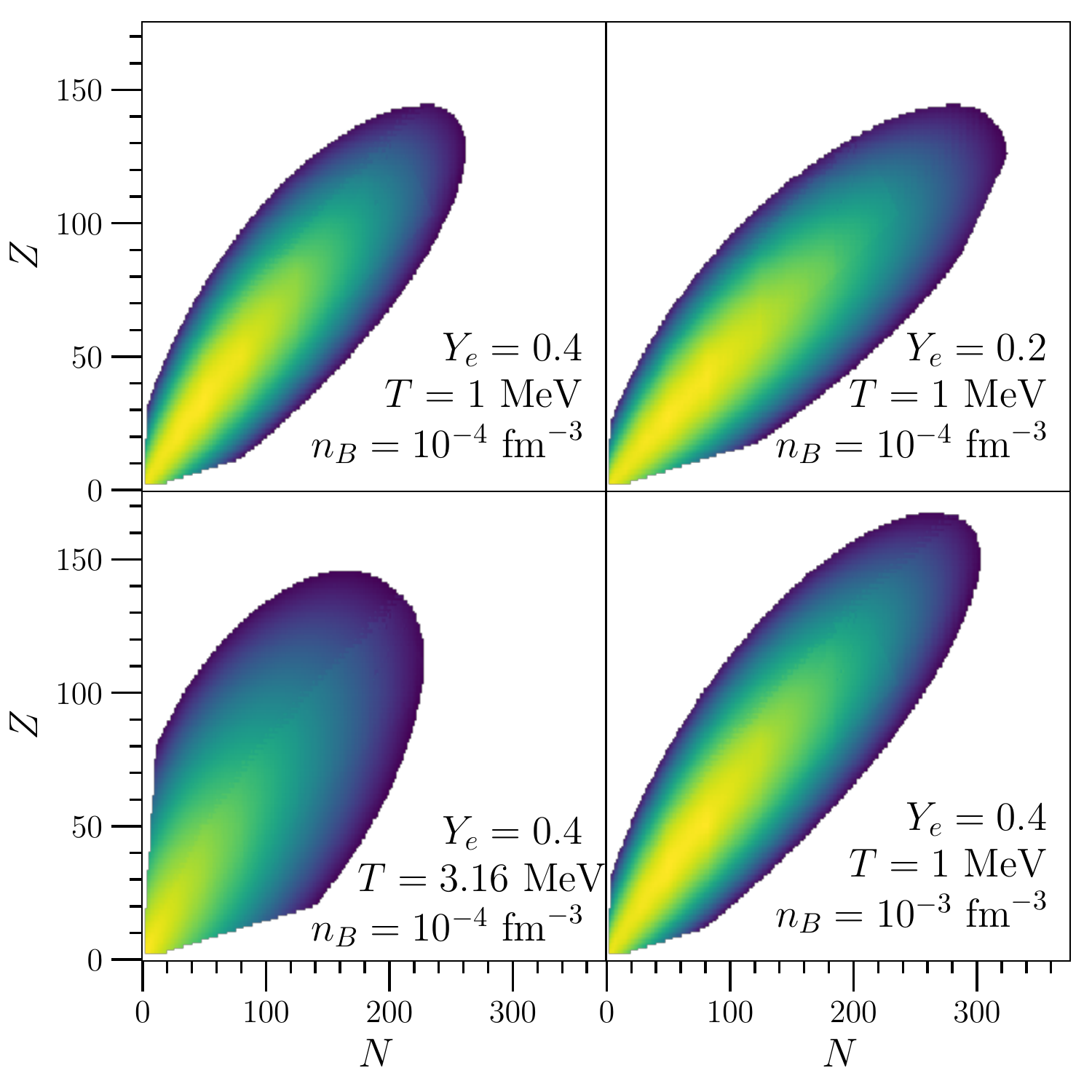}
  \caption{Mass fraction of nuclei in the nuclear chart for matter at
    four selected points, comparable with Ref.~\cite{Shen10}.
  \label{fig:dist}}
\end{figure}

\begin{figure}[h]
  \includegraphics[width=3.1in]{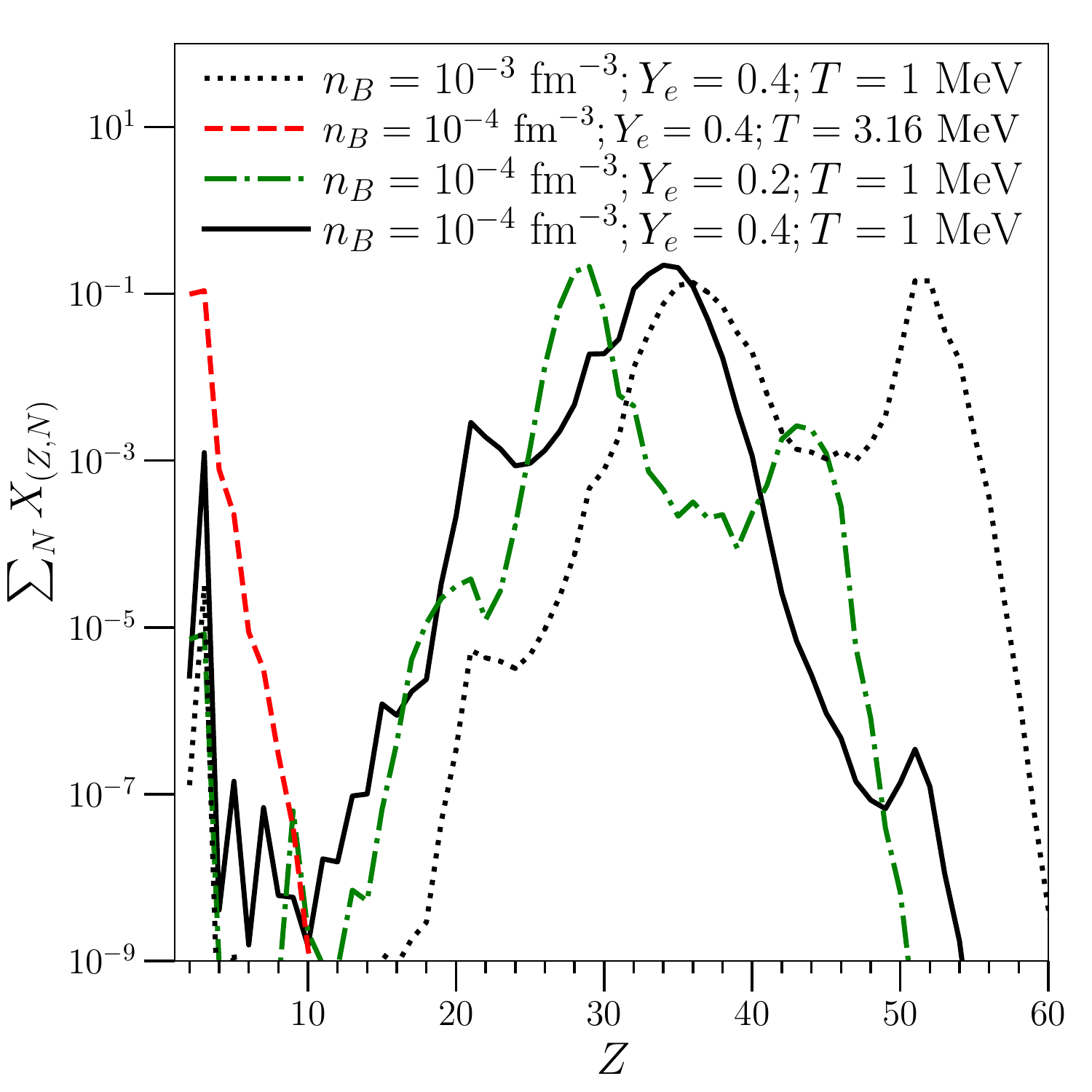}
  \caption{Isotopic distribution for the same four points shown in
    Fig.~\ref{fig:dist}.
  \label{fig:distz}}
\end{figure}

Fig.~\ref{fig:dist} shows the nuclear distribution for selected points in
the EOS as in \cite{Shen10}. Our results are similar, and our
restriction of $Z<7N$ and $N<7Z$ is evident in the linear cutoff in
the distribution near the lower-left corner in each panel.
A significant number of nuclei
participate in the EOS at each point. Even though we do not fully
explore this uncertainty in this work, we find that changing the
distribution can significantly change the transition to nucleonic
matter. This variation may impact core-collapse supernovae and
protoneutron star evolution, as implied by the recent discussion in
Ref.~\cite{Roggero18}. Fig.~\ref{fig:distz} shows the isotopic
distribution for the same four points in the $(n_B,Y_e,T)$ space. The
distribution shows a structure created by the magic numbers (peaks
near Z=28 and Z=50 are evident), as well as a peak at low Z as found
earlier in Ref.~\cite{Souza09b}.


\subsection{Monte Carlo results}

\begin{figure}[h]
  \includegraphics[width=3.1in]{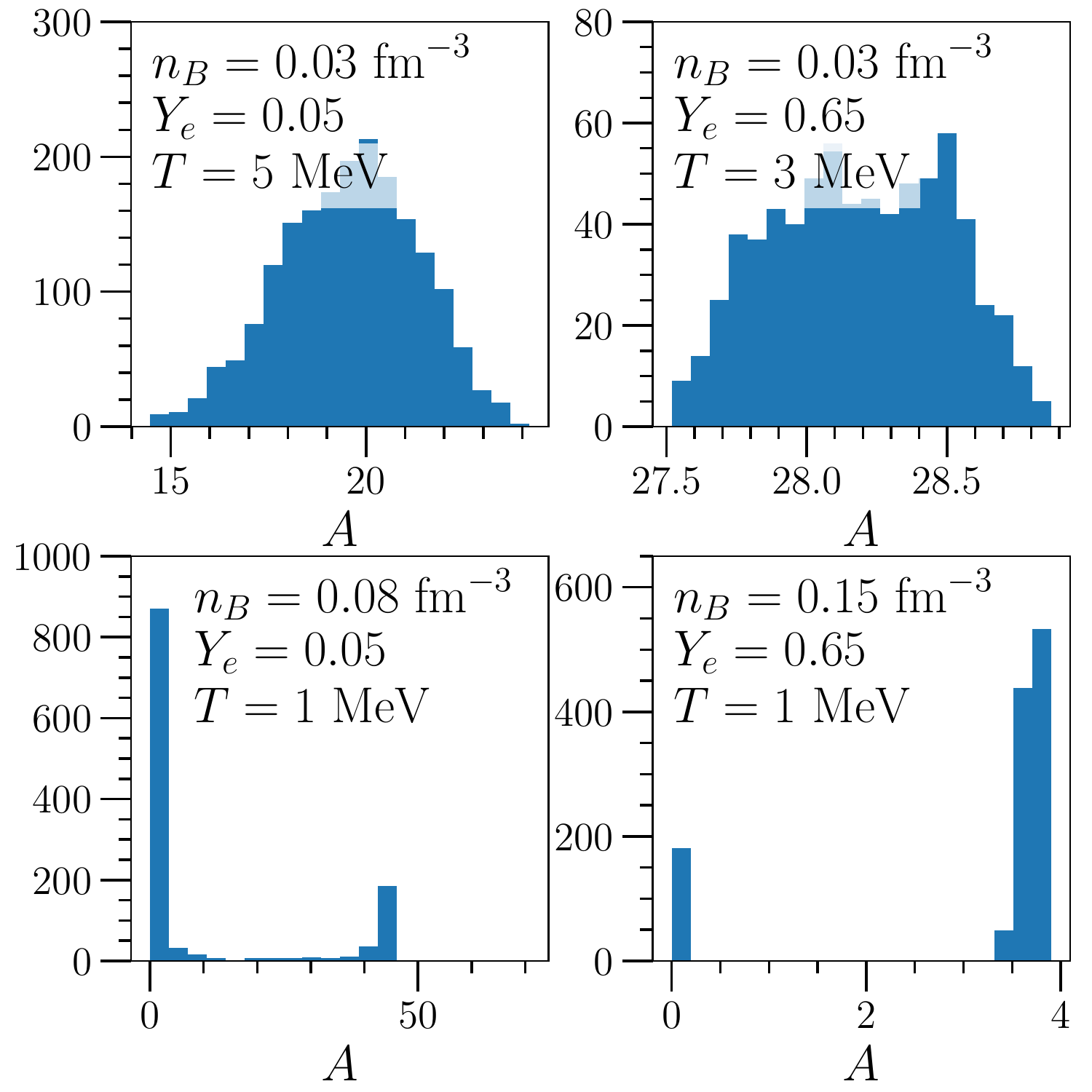}
  \caption{Probability distribution for the average nuclear mass number
    for equations of state generated by our code at four points.
  \label{fig:mcarlo}}
\end{figure}

Fig. \ref{fig:mcarlo} shows four Monte Carlo plots of the average mass
number for some selected points when the seven parameters in our EOS
are randomly selected. The distribution gives uncertainty of the EOS
in subnuclear density at low temperature at four points where the
distribution is nearly maximal. The distribution of A is wider at
extreme values of $Y_e$, the top panels show results for $Y_e=0.05$
and $Y_e=0.65$. The bottom-left panel shows that the probability
distribution is particularly wide for larger densities near the
transition to nucleonic matter in large part because heavy nuclei are
present in some models but not others. This effect persists even
up to large densities, as shown in the lower-right panel, where
nuclei are present for some models but not others.

At some points in the $(n_B,Y_e,T)$ space, the variation shown in
Fig.~\ref{fig:mcarlo} is much smaller than the variation between other
EOS tables. At $n_B=0.03~\mathrm{fm}^{-3}, Y_e=0.05$, and $T=5$ MeV
(corresponding to the upper-left panel of Fig.~\ref{fig:mcarlo}), LS220
gives $A=9$ but STOS gives $A=204$ whereas our result is $19.5 \pm
3.5$. Our variation in some regions, however, is larger than the
variation between EOS tables. At $n_B=0.08~\mathrm{fm}^{-3},
Y_e=0.05$, and $T=1$ MeV (corresponding to the lower-left panel of
Fig.~\ref{fig:mcarlo}), NRAPR gives $A=1$, FSU21 gives $A=0$, FYSS gives
$A=12$, LS220 gives $A=4$, SFHO gives $A=1$, and STOS gives $A=0.15$,
while our result is as large as $A=45$ for some parameterizations.

\section{Discussion}

While we have created a code which can propagate the uncertainties in
the nucleon-nucleon interaction to the resulting equation of state, we
have not yet fully included all of the uncertainties. In particular,
in addition to the several uncertainties which are involved in the
calculation of homogeneous nucleonic matter (discussed in
Ref.~\cite{Du19hd}), there are several additional uncertainties
involving nuclei which we have not included. Pasta structures, which
are present to surprisingly large temperatures, are not included in
the present work. In addition, the modification of the nuclear surface
energy due to the presence of nucleons outside nuclei (see, e.g.,
Refs.~\cite{Lattimer85,Steiner05ia} has not been included in this work.
While these corrections are principally important at lower
temperatures, and are thus subleading, they may impact the
resulting nuclear distribution, particularly in core-collapse
supernovae.

One important consideration is the recent experimental measurement of
a large value for $L$, as measured in PREX-II~\cite{Adhikari21,Reed21}.
While our fiducial model has a smaller value of $L$ one of our
alternate parameterizations has a value of $L=100$ MeV, only 6 MeV
away from the central value suggested in Ref.~\cite{Reed21}.

The nucleon effective mass has been recently shown to be particularly
important for both core-collapse supernovae and
mergers~\cite{Andersen21,Raithel21}. While the parameterizations
tabulated in Table I all use the same Skryme model (which has a
reduced effective mass of 0.904), the zero temperature effective
masses are indeed modified in our full Monte Carlo results presented
in Figure~\ref{fig:mcarlo}. We do not vary the finite-temperature
effective mass from our Skyrme model, SK$\chi$m$^{*}$, because we do
not yet have a probability distribution for the finite temperature
part of the EOS, but this work is in progress. The effective mass,
unlike the equation of state, is not a quantum mechanical observable
(it depends, for example, on the arbitrary demarcation between the
kinetic and potential energy). Thus it only has a unique specification
in the context of a particular model or class of models. However, the
effective mass is important for computing the neutrino mean free path,
which is well-defined, and clearly relevant for simulations of
supernovae and mergers. Thus the best way to properly assess the
impact of the effective mass is construct a probability distribution
of {\em{both}} the equation of state and the neutrino opacities
together. Work on this direction is also in progress.

\section*{Acknowledgements}

The work of XD and AWS was supported by DOE SciDAC grant DE-SC0018232
and the DOE Office of Nuclear Physics. The work of JWH is supported by
the National Science Foundation under Grant No. PHY1652199 and by the
U.S. Department of Energy National Nuclear Security Administration
under Grant No. DE-NA0003841. This research used resources of the
National Energy Research Scientific Computing Center (NERSC), a U.S.
Department of Energy Office of Science User Facility located at
Lawrence Berkeley National Laboratory, operated under Contract No.
DE-AC02-05CH11231. The open-source code for this work,
\href{https://github.com/awsteiner/eos}{https://github.com/awsteiner/eos},
is built upon O$_2$scl~\cite{Steiner14oo}, GSL, HDF5, and
matplotlib~\cite{Hunter:2007}. Tables are available for download at
\href{https://neutronstars.utk.edu/code/eos}
     {https://neutronstars.utk.edu/code/eos}.
  
\bibliographystyle{apsrev}
\bibliography{paper2.bib} 
\end{document}